\documentclass[11pt]{article}
\usepackage{mwe}
\usepackage{graphicx}
\usepackage{subcaption}
\newcommand{\indep}{\rotatebox[origin=c]{90}{$\models$}}
\usepackage{authblk}
\usepackage{subcaption}
\usepackage{blindtext}
\usepackage{graphicx}
\usepackage{amssymb}
\newcommand{\cmark}{\ensuremath{\checkmark}}
\usepackage{multirow}
\usepackage{amsmath}
\usepackage{bm}
\usepackage{amsthm}
\usepackage{float}
\usepackage{cancel}
\usepackage{cite}
\usepackage{comment}
\usepackage{caption}
\usepackage{booktabs,tabularx}
\usepackage{adjustbox}
\usepackage{pgfplotstable}
\newcommand{\subtitle}[1]{%
	\posttitle{%
		\par\end{center}
	\begin{center}\large#1\end{center}
	\vskip0.5em}%
	}
	\date{}
	\usepackage{float}
	\usepackage{multirow}
	\usepackage{pdfpages}
	\usepackage{fullpage}
	\usepackage{verbatim}
	\usepackage{cite}
	\usepackage[english]{babel}
	\usepackage{float}
	\usepackage[hidelinks=true]{hyperref}
	\hypersetup{
colorlinks   = true, 
urlcolor     = blue, 
linkcolor    = blue, 
citecolor    = blue 
}

\usepackage{color}
\usepackage[inner=3cm,outer=2cm,bottom=2cm,top=2cm]{geometry}
\usepackage{setspace}
\usepackage[all]{hypcap}
\usepackage[square,sort,comma,numbers]{natbib}
\usepackage{amsmath,amsfonts}
\usepackage{grffile}
\usepackage[all]{xy}
\usepackage{url}

\newtheorem{thm}{Theorem}

\newtheorem{prop}[thm]{Proposition}

\newtheorem{defn}[thm]{Definition}

\title{Definition, Identification, and Estimation of the Direct and Indirect Number Needed to Treat}
\author{Valentin Vancak$^1$\footnote{Corresponding author: Valentin Vancak, E-mail: valentin.vancak@gmail.com}\, \& Arvid Sjölander$^2$} 

\affil{$^1$Department of Data Science\\ 
Holon Institute of Technology\\
Holon, Israel\\
\vspace{0.3cm} 
$^2$Department of Medical Epidemiology and Biostatistics \\
Karolinska Institutet\\
Stockholm, Sweden}

\begin{document}

\maketitle

\begin{abstract}
	The number needed to treat (NNT) is an efficacy and effect size measure commonly used in epidemiological studies and meta-analyses. The NNT was originally defined as the average number of patients needed to be treated to observe one less adverse effect. In this study, we introduce the novel direct and indirect number needed to treat (DNNT and INNT, respectively). The DNNT and the INNT are efficacy measures defined as the average number of patients that needed to be treated to benefit from the treatment's direct and indirect effects, respectively. We start by formally defining these measures using nested potential outcomes. Next, we formulate the conditions for the identification of the DNNT and INNT, as well as for the direct and indirect number needed to expose (DNNE and INNE, respectively) and the direct and indirect exposure impact number (DEIN and IEIN, respectively) in observational studies. Next,  we present an estimation method with two analytical examples. A corresponding simulation study follows the examples. The simulation study illustrates that the estimators of the novel indices are consistent, and their analytical confidence intervals meet the nominal coverage rates. 
	
	\vspace{0.8cm}
	
\end{abstract}

\section{Introduction}\label{sec:NNT}
The rising prominence of mediation analysis in the last decade underscores its critical role in many research fields, including epidemiology, psychology, and managerial science.\cite{nguyen2021clarifying,  rijnhart2021mediation} This popularity reflects a growing acknowledgment of the need to move from estimating the marginal and conditional total effects and delve into the studied phenomena underlying causal  mechanisms. Mediation analysis, an intrinsically causal concept, has gained popularity for its ability to provide insights into the underlying causal mechanisms through which the exposure affects the outcome. Thus,  mediation analysis answers research questions regarding the nature of the causal mechanism. Particularly, it may answer questions such as: How much (if any) of the exposure effect on the outcome goes through the mediator? However, although many methods have been developed to test the significance of indirect effects,\cite{mackinnon2002comparison} limited attention was devoted to quantifying the absolute and the relative importance of such effects in intuitive units. 

Incorporating effect size measures in mediation analysis is crucial as it addresses the need for a comprehensive and interpretable understanding of the practical importance of observed mediation effects.\cite{preacher2011effect, lange2016commentary} Effect size measures offer a quantitative way to convey the magnitude and direction of the mediated effect. This is crucial for researchers and practitioners seeking to evaluate the mediated pathways' practical relevance and real-world implications. Effect size measures enhance the interpretability of mediation results, allow for meaningful comparisons across studies, and help to bridge the gap between the theoretical and statistical significance and practical relevance of the research in a manner essential for more optimal decision-making and development of effective intervention policies. The most commonly used effect size measure that quantifies the relative importance of the mediation effect is the mediation proportion,\cite{ditlevsen2005mediation} defined as the ratio between the indirect and total effect. However, such a measure has notable limitations.\cite{preacher2011effect} First, the mediation proportion can provide misleading insights into practical significance. For example, a high mediation proportion of a small total effect may be less important in practice than a smaller mediation proportion of a large total effect. Additionally, the mediation proportion can be problematic because it is not always a true proportion. It can be negative or exceed one when the indirect and direct effects are in opposite directions. Another limitation arises when the total effect is small, which can lead to unstable and unreliable estimates. Furthermore, the mediation proportion is scale-dependent, meaning it can produce different effect sizes for the same data when measured on different scales.     Other used effect size measures, e.g., Cohen's~$d$,\cite{preacher2011effect} are unsuitable for mediation analysis since they were originally developed for non-mediated effects.  Therefore, there is a need to construct effect size measures that (1) communicate the results in easily interpretable units, (2) are suitable for diverse applications, and (3) are constructed specifically to quantify indirect and direct effects.

The number needed to treat (NNT) is an efficacy and effect size measure  commonly used in the analysis of randomized controlled trials (RCT), as well as in epidemiology and meta-analyses.\cite{newcombe2012confidence,  vancak2021guide, lee2020cost, verbeek2019cost, da2012methods, mendes2017number}  It is assumed that there are two groups - treated and untreated. In observational studies, the treated group characteristics may substantially differ from the untreated group. Therefore, groupwise efficacy indices were defined.\cite{bender2002calculating} Particularly, for the treated group, the defined groupwise measure is the exposure impact number (EIN), and for the untreated group is the number needed to be exposed (NNE). Each one of these indices answers a unique research question. The populationwise NNT might be of interest if the treatment (exposure) is considered mandatory for the whole population, while NNE and EIN might be of interest in scenarios where the exposure is elective.   The term \textit{treatment} is used interchangeably with the term \textit{exposure} to be consistent with the common terminology in observational studies. The NNE answers questions regarding the impact of exposing the unexposed individuals, and the EIN answers questions regarding the impact of removing the exposure from the exposed individuals. The NNT was originally defined as the average number of patients needed to be treated to observe one less adverse effect.\cite{laupacis1988assessment, kristiansen2002number}  These are two equivalent definitions since avoiding one more adverse effect can be defined as the treatment benefit. The NNE and the EIN were defined analogously as the  average number of patients needed to be exposed to observe one additional exposure benefit on the unexposed and the exposed, respectively. As we are focusing on epidemiological studies, the term \textit{treatment benefit}  is interchangeable with the term \textit{exposure benefit}. The causal meaning is embedded in the definition of the NNE, EIN and NNT.\cite{mueller2022personalized, VancakSjölander+2024} For example, the NNT is usually  defined as a reciprocal of the average exposure effect, which is the reciprocal of the  exposure benefit. Next subsection introduces formal definition these measures. 

\subsection{NNT, NNE and EIN: Notations and Definitions}
Let~$A$ be the exposure indicator where its realization is denoted by the subscript~$a$, such that $a=1$ denotes exposure and $a=0$ non-exposure. Let~$I_1$ be the potential outcome for a given individual if exposed (i.e., if the exposure~$A$ is set to~$1$) and~$I_0$ be the potential outcome for the same individual if the individual was not exposed (i.e., if the exposure~$A$ is set to~$0$). We assume that the outcomes are binary, possibly as a result  from dichotomization of a continuous variable.\cite{vancak2020systematic, kraemer2006size}  A frequent definition of the exposure benefit is   
\begin{align}
	p_t \equiv \mathbb{E}[I_1 - I_0] = \mathbb{P}(I_1 = 1) - \mathbb{P}(I_0=1).	
\end{align}	
This quantity is also known as the average treatment effect (ATE). 
By using the outcome monotonicity assumption, i.e., $ I_1 \ge I_0 $, the ATE is non-negative and $p_t = \mathbb{P}(I_1 = 1, I_0=0)$, thus, the exposure benefit~$p_t$ can be interpreted as probability. Without assuming outcome monotonicity, the target parameter~$p_t$ remains the ATE. However, in such a case it cannot be interpreted in terms of probabilities. 
Notably, even under strict monotonicity, due to sampling variability, the point estimator of the exposure benefit~$p_t$ might still have a negative sign. Since the NNT (NNE, EIN) are defined as the reciprocal values of the exposure benefit, negative~$p_t$ values lead to negative NNT (NNE, EIN), which, in turn,  may lead to difficulties in their interpretation,\cite{grieve2003number, hutton2000number, snapinn2011clinical, sonbo2004cost, kristiansen2002number} bi-modal sample distribution,\cite{grieve2003number} and infinite disjoint confidence intervals~(CIs). Vancak et al.\cite{vancak2020systematic} proposed to resolve the pitfall of singularity at~$0$ and negative signed indices by modifying the original definition of the NNT. The modified NNT is
\begin{align}\label{def.g}
	\text{NNT} \equiv g(p_t) = 
	\begin{cases}
		1/p_t, \quad p_t >0 \\
		\infty ,\quad \,\,\,\, \,  p_t \le 0 \, . 
	\end{cases}
\end{align}   
As such, the best possible NNT value is~$1$, which corresponds to~$p_t = 1$. Namely, a situation where the treatment (exposure) guarantees beneficial outcome.   Whereas, the worst possible NNT value is~$\infty$, which corresponds to~$p_t \le 0$. Namely, a situation where the treatment (exposure) either has no effect or has adverse effects compared to the non-treatment (non-exposure). Notably, the proposed modification helps resolving the practical issues that may arise due to sampling variability. We adopt this modification. Namely, every index (NNT, NNE and EIN, and the forthcoming direct and indirect counterparts) are defined by applying the function~$g$~as in eq.~\eqref{def.g} to the  exposure benefit in the corresponding group. Particularly, the exposure benefit in the $a$th group is defined as
\begin{align}\label{def:p_t(a)}
	p_t(a) \equiv \mathbb{E}[ I_1 - I_0 | A= a], \quad a = 0, 1,
\end{align}
thus, the NNE is defined as $g(p_t(0))$, and the EIN as $g(p_t(1))$.  

  Neither the exposure benefit nor the NNT tell much about the causal mechanisms through which the exposure operates. A possible approach to shed some light on the structure of the causal mechanism is by decomposing the total (causal) effect into two distinct parts - the direct effect and the indirect effect that operates through an intermediate variable called mediator. Such a decomposition propagates into the composition of the NNT itself. Particularly, the NNT is the total-effect efficacy measure. Considering a mediator, one can derive the direct effect NNT that is denoted by DNNT, and the indirect effect NNT, that is denoted by INNT. Such decomposition can be also of interest from a practical point of view. Namely, in situations where the exposure is inevitable, however, the indirect effect accounts for a large proportion of the total effect, one can affect the efficacy of the exposure by intervening on the mediator. Such a motivation can also be valuable for policy makers where the effect of the policy is mediated via an additional variable.

Several studies expressed the need for direct and indirect NNTs (NNEs). For example, a review article on mediators and moderators in early intervention in psychiatry\cite{breitborde2010mediators} expressed the need for such measures to quantify the practical implication of second-generation research in easily interpretable units. In particular, the authors suggest that to better understand the practical significance of a factor mediating the effect of a clinical intervention, one can calculate the NNT at different levels of the mediator (moderator) variable. Namely, efficacy measures like the direct and the indirect NNT may help refine early intervention programs to personalize them and maximize their effectiveness, thus promoting the advancement of the whole research field. Another research that aimed to outline some of the most important methodological challenges in studying acute exacerbation of chronic disease also raised the need for the direct and the indirect NNE.\cite{tsai2009methodological} Particularly, this research emphasizes the importance of treatments' indirect effects estimation for clinical decision making in chronic disease epidemiology and recommend the use of a modified version of the NNE\footnote{Although the authors called it NNT, they applied the calculation on the exposed sub-population. Therefore, it was, in fact, the NNE.} to quantify and communicate the significance of treatments.  A particular subfield of clinical epidemiology that may benefit from introducing indirect NNT and NNEs is allergy research. Since the food allergies are classified as immunoglobulin~E (IgE)-mediated, mixed IgE-mediated, and non-IgE mediated allergies,\cite{du2016prevention} efficacy measures like the indirect and direct NNTs (NNEs) will be naturally suitable for this research subfield.  Allergy research is not an exception but rather a rule in clinical epidemiology. A more general concern frequently considered is that different treatment strategies may have side effects. For example, a highly active antiretroviral therapy for reduction in HIV morbidity and mortality can also affect factors of cardiovascular disease (CVD),\cite{carr2003cardiovascular} which, in turn, affect the morbidity and mortality of the HIV-infected population. For example, if the DNNT is approximately~$2$, while the (total-effect) NNT is~$4$, this suggests that the antiretroviral therapy may have side effects that indirectly increase mortality through CVDs, thereby halving the therapy's effectiveness. In such cases, clinicians might consider addressing CVD-related risk factors to mitigate this indirect pathway and preserve the efficacy of the antiretroviral therapy.  Namely, the indirect effects are frequently inevitable in many clinical situations. Therefore, there is a high demand for measures that can quantify and communicate the effect sizes of the direct, indirect, and total effects separately. The usage of indirect NNTs is not restricted to clinical epidemiology. In sociology and migration studies, several authors\cite{pellegrini2021social}  used a single mediator model to compute path-dependent NNTs to quantify, interpret, and communicate the findings regarding the direct and indirect effects of social exclusion on anti-immigration attitudes. This example demonstrates the usefulness of the indirect NNT (EIN, NNE) not only in clinical epidemiology but also in any research field that analyzes observational data.

The main concern with the aforementioned examples was that even where the various NNTs (NNEs) were computed, it was done in an ad-hoc fashion, thus possibly introducing bias and inconsistency. Moreover, the total, direct, and indirect effect NNTs were not formally defined.  In several studies, the NNT (NNEs) were explicitly causal measures, while it was only implicitly assumed in others.  Therefore, there is a need to define the novel measures formally, and to provide the researchers with the conditions for their identification and consistent estimation. Hence, we start by using Rubin's potential outcomes framework\cite{rubin2005causal} to define the total, direct, and indirect effects indices. The NNE is a one-to-one mapping of the total exposure benefit in the unexposed group. The DNNE and the INNE are one-to-one mappings of the direct and indirect exposure benefits, respectively, in the  the same group.  Therefore, first, we decompose the exposure benefit into direct and indirect effects (benefits). Next, we decompose the indirect effect path into two parts: the exposure-mediator and the mediator-outcome. Next,  we formulate the conditions for  identification and consistent estimation  of the exposure benefits and the corresponding indices.  We present two analytical examples and a corresponding simulation study. The simulation study illustrates that the novel estimators are statistically consistent, and their confidence intervals meet the nominal coverage rates. 

\subsection{Direct and Indirect NNT, NNE and EIN: Notations and Definitions}\label{subsec:decomp_nnt}
  Let $M$ be the mediator.  We assume that the mediator~$M$ is a binary variable. Notably, the presented methodology can be readily extended to non-binary exposures and mediators. However, for the clarity of exposition and subsequent computations, we adhere to the binary case. Let $I_{a,m}$ be the potential outcome of~$I$ when~$A$ and~$M$ are set to~$a \in \{0, 1\}$ and~$m \in \{0, 1\}$, respectively. Let $M_{a'}$ be the potential outcome of~$M$ when~$A$ is set to $a' \in \{0, 1\}$. Thus, $I_{a,M_{a'}}$ be the potential outcome of~$I$ when~$A$ is set to~$a$, and~$M$ is what it would have been if~$A$ is set to~$a'$, where~$a$ may differ from~$a'$. The quantity~$I_{a,M_{a'}}$ is referred to as nested counterfactual since the potential outcome of~$M$ (when $A$ is set to $a'$) is nested in the  potential outcome of~$I$.    Next we define the main effects of interest and the corresponding indices; the Total Effect and the corresponding total effect indices (NNT, NNE, EIN), the Natural Direct Effect and the corresponding direct effect indices (DNNT, DNNE, DEIN), and the Natural Indirect Effect with the corresponding indirect effect indices (INNT, INNE, IEIN).

\begin{defn}[Corollary of the modified NNT (NNE, EIN) definition in eq.~\eqref{def.g} and eq.~\eqref{def:p_t(a)}]\label{defn:TNNT} The total effect (TE) in the $a$th group, $a\in \{0,1\}$, is defined as
	$
	p_t(a) = \mathbb{E}[I_{1} - I_{0}|A=a] = \mathbb{E}[I_{1,M_1} - I_{0,M_0}|A=a],
	$
	therefore, the NNE and the EIN are defined as 
	\begin{align}
		\mathrm{NNE} \equiv g(p_t(0)), \quad  \mathrm{EIN} \equiv g(p_t(1)).
	\end{align}  The marginal	total effect  is defined as 
	$
	 p_t =  \mathbb{E}[I_1 - I_0] = \mathbb{E}[p_t(A)].
$
Therefore, the total effect NNT  is 
\begin{align}
\mathrm{NNT} \equiv g(p_t).	
\end{align}
\end{defn}
 In the definition of the total effect~$p_t$, we ignore the mediator when contrasting between the exposed and the unexposed groups. Ignoring the mediator is equivalent to letting the mediator~$M$ attain the value it would have attained where the exposure~$A$ is set to~$a$, i.e.,~$M_a$. The total effect NNT (NNE, EIN)  is a marginal effect index that encapsulates the effects of all the causal paths between the exposure and the outcome. Therefore, it is a crude efficacy measure since it makes no distinction between the direct and the indirect causal paths. Specifically, if the direct and the indirect effects are in opposite directions, the total effect NNT conveys the net exposure effect that can differ substantially from the direct effect NNT which conveys the non-mediated exposure effect.  Next, we define the natural direct and the natural indirect effects. The term natural means that we don't set the mediator~$M$ to the same value for all subjects, however,  we let it attain whatever value it would have been after setting the exposure~$A$ to a certain value~$a$. Namely, we, hypothetically, intervene on the exposure, but do not have any further control over the mediator.
\begin{defn}[]\label{defn:DNNT}
The Natural Direct Effect   in the $a$th group, NDE$(a)$. $a\in \{0,1\}$, is defined as
$
p_d(a) = \mathbb{E}[I_{1,M_{1}} - I_{0, M_{1}}|A=a] ,
$
therefore, the direct effect NNE (DNNE) and EIN (DEIN) are defined as
\begin{align}
	\mathrm{DNNE} \equiv g(p_d(0)), \quad  \mathrm{DEIN} \equiv g(p_d(1)).
\end{align} 	
The  marginal direct effect is defined as 
$
p_d = \mathbb{E}[I_{1,M_{1}} - I_{0, M_{1}}] = \mathbb{E}[p_d(A)].
$ Therefore, the direct effect NNT, DNNT, is defined as 
\begin{align}
	\mathrm{DNNT} \equiv g(p_d).
\end{align}
\end{defn}
   The direct effect measures defined as a function of exposure effect that is obtained by contrasting exposure with non-exposure for the main potential outcome of~$I$ while holding the exposure constant at~$1$ for the nested potential value of the mediator~$M$. The direct effect NNT is the average number of patients that need to be treated (exposed) in order to observe one additional benefit that is caused directly by the treatment (exposure) while the mediator is kept at the value that it would have attained had the patient been exposed. Namely, the DNNT quantifies the size of the non-mediated effect of the treatment (exposure) on the outcome.\footnote{Notably, in the mediation analysis literature, the well-known controlled direct effect (CDE) is defined as $p_c(m; a) = \mathbb{E}[I_{1,m} - I_{0,m} \mid A = a]$, representing the effect of setting the mediator to a fixed value~$m$ in the $a$th exposure group.\cite{pearl2022direct} Based on this definition, one can construct both marginal and conditional (groupwise) controlled direct effect NNT (NNE, EIN). However, since the $g\left( p_c(m;a)\right)$ is just the adjusted NNT\cite{vancak2022number} for $M=m$ it will not be discussed any further in this article. For more details on the equivalence of $g\left( p_c(m; a)\right)$ and the adjusted NNT, please refer to Appendix~\ref{app:CDE}}

\begin{defn} \label{defn:INNT}
The  Natural Indirect Effect (NIE) in the $a$th group, $a\in \{0,1\}$, is defined as
$
p_i(a) = \mathbb{E}[I_{0,M_{1}} - I_{0, M_{0}}|A=a],
$
therefore, the indirect effect NNE (INNE) and EIN (IEIN) are defined as
\begin{align}
	\mathrm{INNE} \equiv g(p_i(0)), \quad  \mathrm{IEIN} \equiv g(p_i(1)).
\end{align} 
The marginal indirect effect is defined as $p_i = \mathbb{E}[I_{0,M_{1}} - I_{0, M_{0}}]  = \mathbb{E}[p_i(A)]$. Therefore, the indirect effect NNT, INNT, is defined as 
\begin{align}
\mathrm{INNT} \equiv  g(p_i).	
\end{align}
\end{defn}
Namely, the INNT is defined as a function of the exposure effect that is obtained by contrasting exposure with non-exposure for the nested potential outcome of the mediator~$M_a$ while setting the exposure constant at~$0$ for the main potential outcome of~$I$. The interpretation of the INNT is the average number of patients that need to be treated to observe one additional benefit that is caused solely by the mediated effect of the exposure on the outcome.
Namely, the INNT conveys the size of the effect on the outcome that happens specifically through the pathway involving the mediator, isolating this part from any direct effect the exposure may have on the outcome. The total effect~$p_t$ can be decomposed into addition of the two natural effects. Namely, the total effect in the $a$th group can be written as a sum of the $p_i(a)$ and the $p_d(a)$\cite{pearl2012causal}, i.e., 
\begin{align}\label{prod_dir_plus_indir}
	p_t(a) = p_d(a) + p_i(a), \quad a = 0, 1. 
\end{align}
Clearly the additive decomposition doesn't hold for the NNT since~$g$ as defined in eq.~\eqref{def.g} is non-linear. 

\subsection{Factorization of the direct and indirect effects}\label{subsec:factor}

We adopt two sufficient, but not necessary, structural assumptions to simplify interpretation and align with familiar linear-model intuition. For the NIE, we assume effect homogeneity across principal strata\cite{forastiere2018principal}, i.e., 
$
\mathbb{E}\big[I_{0,1}-I_{0,0}\mid M_0, M_1\big]
= \mathbb{E}\big[I_{0,1}-I_{0,0}\big]
$, and for the NDE we assume principal-strata invariance of the controlled direct effect, i.e., $
\mathbb{E}[I_{1,m}-I_{0,m}\mid A=a, M_a]
= \mathbb{E}[I_{1,m}-I_{0,m}\mid A=a],
$
for $a, m \in \{0, 1\}$. These assumptions yield closed-form factorizations presented in Propositions~\ref{thm:NIE} and~\ref{thm:NDE}. Particularly,  the NIE becomes a product of an exposure-mediator contrast and a mediator-outcome controlled contrast, while the NDE becomes a weighted sum of controlled direct effects with weights given by the mediator prevalence under exposure. These forms mirror the no-interaction linear model,\cite{robins1992identifiability, pearl2012causal} make each pathway's contribution transparent and modular, and preserve the intended interpretation of the indices as contrasts between exposure arms rather than across mediator strata. We emphasize that the assumptions are invoked only to allow the algebraic factorization that streamlines exposition and simplifies the subsequent numerical example. Notably, identification of the indirect and direct effects requires a different set of assumptions presented in Section~\ref{sec:ident} and is independent of the factorization presented below.

\begin{prop}\label{thm:NIE} 
	Let $I, A, M$ be  binary outcome, exposure, and mediator, respectively. Assuming effect homogeneity across principal strata, the NIE, denoted by~$p_i$, satisfies
	\begin{align}\label{binary_mediation}
		p_i = 
		\mathbb{E}\left[M_1 - M_0 \right ] \mathbb{E}\left[I_{0, 1} - I_{0, 0} \right].
	\end{align}
	Analogously, for the subgroup with \( A = a \), the conditional NIE, denoted by $p_i(a)$, is given by 
	\begin{align}\label{def:NIE_BINARY}
		p_i(a) = 
		\mathbb{E}\left[M_1 - M_0 \mid A=a \right]  \mathbb{E}\left[I_{0, 1} - I_{0, 0} \mid A=a \right], \quad a =0, 1.
	\end{align}
	See Appendix~\ref{app:proof_NIE} for the full proof.
\end{prop}

\begin{prop}\label{thm:NDE} 
		Let $I, A, M$ be  binary outcome, exposure, and mediator, respectively. Assuming  principal–strata invariance of the controlled direct effect at each mediator level, the NDE, denoted by \( p_d \), satisfies
	\begin{align}\label{binary_direct}
		p_d = \mathbb{E}\left[I_{1, 0} - I_{0, 0} \right] (1 - \mathbb{E}[M_1]) + \mathbb{E}\left[I_{1, 1} - I_{0, 1} \right] \mathbb{E}[M_1].
	\end{align}
	Analogously, the conditional NDE for the group with \( A = a \), denoted by \( p_d(a) \), is given by
	\begin{align}\label{def:NDE_BINARY}
		p_d(a) = \mathbb{E}\left[I_{1, 0} - I_{0, 0} \mid A = a \right]  (1 - \mathbb{E}[M_1 \mid A = a]) + \mathbb{E}\left[I_{1, 1} - I_{0, 1} \mid A = a \right] \mathbb{E}[M_1 \mid A = a],
	\end{align}
	for  \( a \in \{0, 1\} \). See Appendix~\ref{app:proof_NDE} for more details.
\end{prop}

\section{Numerical example}\label{sec: num_example}
Using the definitions in subsection~\ref{subsec:decomp_nnt} and the factorizations in subsection~\ref{subsec:factor}, we present an illustrative numerical example of the novel measures in the context of migration economics. Emigration of highly educated individuals is an issue that concerns many developing countries. This phenomenon is also known as a ``brain drain."\cite{docquier2012globalization} Brain drain, which has a direct negative effect on the sending countries as it decreases the number of highly skilled individuals in the labour force, may ultimately contribute to human capital formation by an indirect effect that such emigration induces via younger population education patterns.  Specifically, countries with relatively low levels of human capital and low emigration rates experience a total beneficial effect of brain drain. The essence of the argument is that since the return to education is higher abroad, migration prospects can raise the expected return to human capital and encourage more people to invest in education in their home country.\cite{beine2001brain} Namely, although a brain drain has an intrinsically negative direct effect on the home country's economy, its positive indirect effect may mitigate the negative direct effect and even exceed it.  Beine et al.~\cite{beine2008brain} proposed a probabilistic model with counterfactual reasoning in order to analyze the aforementioned phenomena. Policymakers, both in the absorbing countries and the sending countries, may be interested in the mechanism through which higher education encourages the emigration of highly skilled workers. If the effect of education is mediated (moderated) by another variable, the policymaker may desire to intervene on this variable in order to encourage (or discourage) emigration incentives. Therefore, testing for the presence of a mediator and determining its relative effect size may be an important step toward formatting effective intervention strategies and forming immigration (emigration) policies. Drawing from this example, we present a simple probabilistic model that illustrates the usefulness of the novel measures in such a context. The numbers presented in the following model are chosen for illustrative purposes and may not resemble the migration patterns in any specific country.         
 
  Let the exposure~$A$ be a binary indicator of possessing undergraduate academic degree, the outcome~$I$ is indicator of emigration, and the mediator~$M$ is an indicator of a command of a foreign language. The idea is that the skills and the professional knowledge acquired in undergraduate degrees may encourage emigration (which is the direct effect) to another country in pursue of better employment opportunities. However, in most undergraduate degree programs, there are also mandatory foreign language studies that are not necessarily related to the main course of studies and the acquired profession. A command of a foreign language may also by itself encourage emigration.

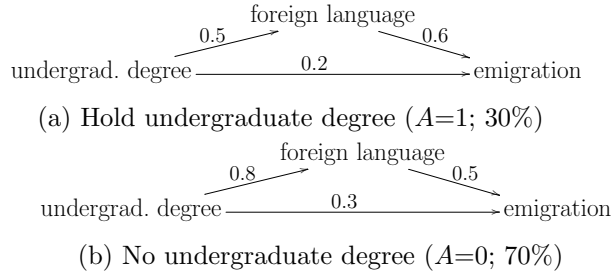
\begin{figure}[ht]
	\centering
	
	\begin{subfigure}[t]{0.7\textwidth}
		\centering
		\scalebox{0.55}{%
			\xymatrixrowsep{0.6cm}
			\xymatrixcolsep{1.2cm}
			\xymatrix{
				& {\LARGE \text{foreign language}} \ar[rd]^{\text{\Large 0.6}} & \\
				{\LARGE \text{undergrad.\ degree}} \ar[ru]^{\text{\Large 0.5}}
				\ar[rr]^{\text{\Large 0.2}} &&
				{\LARGE \text{emigration}}
			}
		}
		\caption{\small Hold undergraduate degree ($A{=}1;\,30\%$)}
		\label{fig:dag_degree_yes}
	\end{subfigure}
	\hspace{0.04\textwidth}      
	
	\begin{subfigure}[t]{0.7\textwidth}
		\centering
		\scalebox{0.55}{%
			\xymatrixrowsep{0.6cm}
			\xymatrixcolsep{1.2cm}
			\xymatrix{
				& {\LARGE \text{foreign language}} \ar[rd]^{\text{\Large 0.5}} & \\
				{\LARGE \text{undergrad.\ degree}} \ar[ru]^{\text{\Large 0.8}}
				\ar[rr]^{\text{\Large 0.3}} &&
				{\LARGE \text{emigration}}
			}
		}
		\caption{\small No undergraduate degree ($A{=}0;\,70\%$)}
		\label{fig:dag_degree_no}
	\end{subfigure}
	
	\caption{\small Directed acyclic graphs (DAGs) representing the illustrated structural model
		under two exposure conditions.  Figure (a), $A = 1$, depicts
		individuals who hold an undergraduate degree. Figure (b),
		$A = 0$, depicts individuals who do not hold an undergraduate degree.  The binary outcome~$I$ is emigration
		status, influenced either directly by educational attainment~$A$ or
		indirectly via foreign-language proficiency~$M$.  Numbers on the arrows are
		the assumed causal effects.}
	\label{fig:dag_dual_migration}
\end{figure}

Assume the following quantities: For the group that holds, in fact, an undergraduate degree, i.e., $A=1$, the direct effect as defined in eq.~\eqref{binary_direct} is $p_d(1)  = 0.2$. For the indirect effect, we need two additional quantities: The effect of undergraduate education on command of a foreign language, and the effect of a foreign language command on emigration (for the group which members hold an undergraduate degree). Assume, for example, that~$50\%$ of those who hold an undergraduate degree acquire a command of a foreign language as a result of the academic studies, i.e., $\mathbb{E}[M_1 - M_0|A=1] = 0.5$.  Additionally, assume that~$60\%$ of this group members will emigrate solely because of the command of foreign language, i.e., $\mathbb{E}[I_{0,1} - I_{0, 0}|A=1] = 0.6$.  Therefore, the indirect effect, as defined in eq.~\eqref{def:NIE_BINARY}, is $p_i(1) = 0.5 \times 0.6 = 0.3$. Hence, the total effect for the exposed, as defined in eq.~\eqref{prod_dir_plus_indir}, is
\begin{align}
	p_t(1) = 0.2 + 0.3  = 0.5. 
\end{align}    
Therefore, using the definitions in~\eqref{defn:TNNT}, \eqref{defn:DNNT}, and~\eqref{defn:INNT}, the EIN$ = 1/0.5 = 2$, DEIN$ = 1/ 0.2 = 5$, and IEIN$ = 1/0.3 = 3\frac{1}{3}$. For the unexposed group, i.e.,  those who do not hold an undergraduate degree, i.e., $A=0$, assume the following quantities: The direct effect probability, as defined in eq.~\eqref{binary_direct}, is $p_d(0) = 0.3$, the indirect effect probability is $p_i(0) = 0.8 \times 0.5 = 0.4$, where $\mathbb{E}[M_1 - M_0|A=0] = 0.8$, and   $\mathbb{E}[I_{0,1} - I_{0, 0}|A=0] = 0.5$. Hence, the total effect for the unexposed, as defined in eq.~\eqref{prod_dir_plus_indir}, is
 \begin{align}
 	p_t(0) = 0.3 + 0.4  = 0.7. 
 \end{align}
 The DAG in Figure~\ref{fig:dag_dual_migration} provides graphical illustration of the structural model of the aforementioned example and the corresponding quantities.
 
Therefore, using the definitions in~\eqref{defn:TNNT}, \eqref{defn:DNNT}, and~\eqref{defn:INNT}, the NNE$ = 1/0.7 = 1.43$, DNNE$ = 1/ 0.3 = 3\frac{1}{3}$, and INNE$ = 1/0.4 = 2.5$.
Assume, for example, that~$30\%$ of the population over the age of~$25$ holds (at least) an undergraduate degree, thus the populationwise total, direct, and indirect effects, as defined in~\eqref{defn:TNNT}, \eqref{defn:DNNT}, and~\eqref{defn:INNT}, respectively, are the following weighted means 
\begin{align*}
p_t &= 0.5 \times 3/10 + 0.7 \times 7/ 10 = 	0.64\\
p_d & = 0.2 \times 3/10 + 0.3 \times 7/ 10 = 	0.27\\
p_i & = 0.3 \times 3/10 + 0.4 \times 7/ 10 = 	0.37,
\end{align*}
respectively. Therefore, the NNT$ = 1/0.64 = 1.56$, the DNNT$ = 1/0.27 = 3.7$, and the INNT$ = 1/0.37 = 2.7$. This example demonstrates that the probability of different groups to emigrate is affected differently by undergraduate education directly and indirectly via command of a foreign language acquired during undergraduate studies. Hence, the policymakers may construct very different policies as a function of what they would like to achieve. For example, for the unexposed group, i.e., those who do not hold an undergraduate degree, the effect of command of a foreign language is very high (INNE $= 2.5$). Namely, among those with no undergraduate education, we need to raise the foreign language skills to $2.5$ individuals to a level it would have been if these people had taken an undergraduate education in order to force one more individual to emigrate.
In contrast, the size of the direct effect of the academic degree on emigration probability is smaller (DNNE $= 3\frac{1}{3}$) compared to the size of the indirect effect. Namely,  among those with no undergraduate education but with foreign language command at the same level that it would have attained in academic education,  $3\frac{1}{3}$ individuals need to obtain undergraduate education in order to force one more individual to emigrates.  Therefore, if the policymaker is interested in encouraging emigration, he can offer foreign-language courses that aim to provide foreign-language command at the same level that would be attained in undergraduate studies.  On the other hand, for the exposed group (which studies undergraduate degrees), a policy that strengthens foreign language education as a part of undergraduate studies by offering subject-matter courses in foreign languages might constitute a more effective intervention since it may enhance the indirect effect of the undergraduate degree studies on the emigration probability. This example illustrates that the novel measures may help to construct different policies for different sub-populations as a function of their unique characteristics and the strength (size) of the mediated effect.

\section{Identification}\label{sec:ident}

Since the novel indices are deterministic, monotone functions of the group-specific and marginal NIE and NDE, their identification requires no additional assumptions beyond those used to identify the underlying NIE and NDE. The fundamental problem of causal inference is that we cannot observe both \(I_1\) and \(I_0\) for the same individual \cite{holland1986statistics}, because an individual is either exposed or unexposed. We assume consistency,\cite{rubin1974estimating} namely, if an individual actually receives~$A=a$ and~$M=m$, then its observed mediator and outcome equal the corresponding potential values, i.e., $M = M_a$, and $I = I_{a,m}$. In practice this requires well-defined, correctly measured interventions and no interference.\cite{ pearl2010consistency}
Identification of contrasts that are functions of \(I_0\) and \(I_1\) requires additional assumptions such as positivity.\cite{petersen2012diagnosing} Namely, for all \(l\) in the support of \(L\),
\[
0<\mathbb{P}(A=a \mid L=l)<1, \ \ a\in\{0,1\}, 
\qquad
\mathbb{P}(M=m \mid A=a, L=l)>0, \ \ a, m = 0, 1.
\]
 Positivity means that for every covariate pattern~$L=l$ observed in the data, both exposure levels occur with nonzero probability, i.e., there are some treated and some untreated units, so identification does not rely on extrapolation. However, the problem of identification  is  amplified  by introducing the nested counterfactuals where certain quantities are not observable by their definition. For example,  the nested counterfactual $I_{a, M_{a'}}$ where $a \neq a'$, are defined by the so-called ``cross-world" interventions, and thus they are not observable even in RCTs where both the exposure and the mediator are manipulable. Yet, since the novel direct and indirect NNT (NNE, EIN) are functions of contrasts that involve this type of counterfactuals, their identification depend on the identification of nested counterfactuals. To ensure it,  we need to assume several more assumptions. Particularly, we require that no unmeasured confounders affect the outcome~$I$, the mediator~$M$, and the exposure~$A$, or any pair of these variables, conditioned on~$L$ which is the set of measured confounders. Formally, such assumptions can be summarized by two conditional independence statements. The first are known as sequential ignorability:\cite{imai2010identification, imai2010general}

\begin{align}\label{def:seq_ign}
	\{I_{a', m}, M_a\} \indep A \mid  L, \quad   I_{a', m} \indep M_a \mid A=a, L, \quad \text{for all } a, a', m = 0, 1.
\end{align}
 The first assumption on the left-hand side is  ignorability of the treatment, and it states that conditioned on the observed confounders~$L$,  the exposure is independent of the potential outcomes and the potential mediators. Namely, no unmeasured confounders confound the effect of the exposure and/or the mediator on the outcome. Practically, it means that after adjusting for the measured covariates~$L$
 (e.g., age, baseline measurements), the decision to treat (expose) behaves “as if randomized” with respect to the potential mediator and outcome. Such an assumption holds by design in a randomized controlled trial, either unconditionally under simple randomization, or conditional on~$L$ under stratification covariates.
  The second ignorability assumption, on the right-hand side, is mediator ignorability, and it  states that conditioned on the observed confounders~$L$ and the exposure group $A=a$, the potential mediator and the potential outcomes are independent. Namely, no unmeasured confounders confound the outcome and the mediator relation. Practically, within each treatment arm and after adjusting for the measured covariates~$L$, individuals who differ in the mediator value are comparable with respect to the outcome  Such an assumption is not guaranteed even in RCTs where the exposure (treatment) is randomized but not the mediator. Such assumption depends on the subject-matter knowledge and generally untestable with observed data alone. Please refer to directed acyclic graphs (DAGs) of twin causal networks in Figures~\ref{dag:twin_med1},~\ref{dag:twin_med2}, and~\ref{dag:twin_med3} in the Appendix for a graphical illustration of the sequential ignorability assumptions.   If these assumptions are satisfied and $L=\emptyset$, we can obtain an identification formula of the NIE for discrete mediator~$M$ and binary exposure~$A$ using the mediation formula.\cite{pearl2012causal} 
This  formula represents the average change in the outcome~$I$ caused by the exposure~$A$ (vs. non-exposure) after deduction of the direct effect of~$A$ on~$I$.  For a binary mediator~$M$, the mediation formula boils down to the observed equivalent of eq.~\eqref{def:NIE_BINARY}
\begin{align}\label{indirect_effect_obs}
	p_i  = \left(   \mathbb E[M \mid  A=1] - \mathbb E[ M \mid A=0]       \right)\left( \mathbb E[I\mid A = 0, M=1] - \mathbb E[I \mid A = 0, M=0]    \right) .
\end{align}
The quantities $\mathbb E [ M\mid A=a]$ and $\mathbb E[I\mid A=0, M=m]$ can be estimated by using sample frequencies. A scenario that satisfies the sequential ignorability and $L=\emptyset$ is identical to randomized allocation to exposure and non-exposure, and that the mediator is unconfounded with the outcome. In such scenario,  there is no systematic difference between the exposed group~$A=1$ and the unexposed~$A=0$. Therefore, the groupwise conditional indices equal each other and equal to the corresponding marginal index. Namely,  NNT$=$EIN$=$NNE, DNNT$=$DEIN$=$DNNE, and INNT$=$IEIN$=$INNE. As such, it is sufficient to compute the NNT triple (NNT, DNNT, INNT) to summarize the information of interest regarding the exposure efficacy. However,  in observational studies, where the allocation to exposure is not randomized, each arm has its own groupwise direct and indirect index that differs from each other and from the marginal indices. Assuming that the measured covariates~$L$ are sufficient for  confounding control is essential for the identification of the novel direct and indirect indices.
The control for confounding includes the confounders of the exposure-mediator, exposure-outcome and mediator-outcome effects. 
Alternatively to mediator ignorability, we can assume principal ignorability,\cite{forastiere2018principal} namely, 
\begin{align}\label{def:princ_ignor}
	I_{a,m} \indep (M_0, M_1) \mid L, \quad  \text{for all } a, m = 0, 1.
\end{align}
Namely, we assume that the distribution of the potential outcome $I_{a, m}$ is the same across principal strata, condtioned on~$L$.\cite{frangakis2002principal} Principal stratification with respect to a post-treatment variable~$M$ partitions units into latent strata indexed by their joint potential mediator values under exposure and unexposure, i.e., $(M_0, M_1)$. Such an assumption is required for identification of principal causal effects, i.e., effects defined conditioned on a principal strata.\cite{ding2017principal} Practically, it means that after adjusting for the measured covariates~$L$, an individual's  latent mediator type $(M_0, M_1)$ 
 does not predict what their outcome would be under a given treatment and mediator setting. Therefore, individuals from different principal strata are comparable for the outcome.   Notably, principal strata ignorability~\eqref{def:princ_ignor} implies ignorability of the mediator, therefore, by assuming the former, we don't need to state the latter explicitly. Moreover, principal ignorability assumption might be more easier to justify since thinking in terms of homogeneity across principal strata may be more intuitive than in terms of no unmeasured confounding of the mediator-outcome relationship.\cite{forastiere2018principal} Additionally, if we will assume monotonicity of the mediator, i.e., $M_1 \ge M_0$. Namely, no negative effects of the treatment on the outcome, then, we can drop principal ignorability and keep only the sequential ignorability in order to identify the novel indices. Notably, such an assumption is frequently holds by design in clinical trials and sociological studies such as presented in Section~\ref{sec: num_example}. Table~\ref{tab:nie-id-assumptions} summarizes the assumptions needed for identification of the novel indices.


\begin{table}[htbp]
	\centering
	\scriptsize
	\setlength{\tabcolsep}{1pt}
	\renewcommand{\arraystretch}{1.3}
	\caption{Assumptions needed for identification of the indirect and direct NNT, NNE, and EIN}
	\label{tab:nie-id-assumptions}
	\begin{tabular}{lcccc}
		\toprule
	\textbf{Assumption sets} &\textbf{ Treatment ignorability} & \textbf{ Mediator ignorability} & \textbf{ Principal ignorability} & \textbf{ Mediator monotonicity} \\
		 & $(I_{a,m},\,M_a)\indep A \mid L$ & $I_{a,m}\indep M_a\mid A, L$ & $I_{a,m}\indep(M_0,M_1) \mid L$ & $M_1 \ge M_0 $ \\
		\midrule
	 1 &	\cmark &  & \cmark &  \\
		\midrule 
	2 &	\cmark & \cmark &  & \cmark \\
		\bottomrule
	\end{tabular}
	
	\vspace{0.5ex}
	\begin{minipage}{0.98\linewidth}\footnotesize
We assume a binary exposure 
$A$
and mediator 
$M$. The outcome is 
$I$, and 
$L$ denotes confounders. Each row lists an alternative set of assumptions sufficient for identification of the direct and indirect NNT, NNE, and EIN. For all sets, we additionally assume consistency and positivity. 
	\end{minipage}
\end{table}

As discussed in Subsection~1.1, if we additionally assume monotonicity of the outcome with respect to the treatment (\(I_{1}\ge I_{0}\); no harm from exposure), the total-effect~$p_t$ admits a probabilistic interpretation. If, moreover, we assume monotonicity of the controlled outcome with respect to the mediator, i.e.,  \(I_{0,1}\ge I_{0,0}\) together with mediator monotonicity, the indirect effect~$p_i$  can likewise be interpreted as a probability. Finally, if we also impose controlled-outcome monotonicity, i.e.,  \(I_{1,m}\ge I_{0,m}\), for \(m\in\{0,1\}\), the natural direct effect~$p_d$ admits the same probabilistic reading. Importantly, the monotonicity assumptions are used solely to simplify interpretation and are not required to define or identify any effect or its corresponding indices.

Treatment and principal ignorability  together with consistency \cite{cole2009consistency} identifies the expectations appearing in Eqs.~\eqref{binary_mediation} and~\eqref{def:NIE_BINARY}. Alternatively, the same quantities are identified under sequential ignorability and  mediator monotonicity. Specifically, for the potential mediator we have
\begin{align}
\mathbb{E}[M_a] = \mathbb{E}[ \mathbb{E}[M_a| L] ] = \mathbb{E}[ \mathbb{E}[M_a| A=a, L] ] = \mathbb{E}[ \mathbb{E}[M|A=a, L] ] , 	 \quad a  = 0, 1,
\end{align}
and for the potential outcome we have
\begin{align}
	\mathbb{E}[I_{a,m}] = \mathbb{E}[\mathbb{E}[I_{a,m}|A=a, M=m, L]] 
	= \mathbb{E}[\mathbb{E}[I|A=a, M=m, L]], \quad a  = 0, 1.
\end{align}
Analogously, the  expected unobserved potential outcomes and mediators of the groupwise indirect effect~$p_i(a)$ in equations~\eqref{prod_dir_plus_indir} and~\eqref{def:NIE_BINARY} can also be identified using either set of assumptions.\cite{sjolander2018estimation} Namely, for the potential mediator we have 
\begin{align}\label{identification_counter_med}
	\mathbb{E}[M_a|A=1-a] = \mathbb{E}[\mathbb{E}[M|A=a, L]|A=1-a], \quad a  = 0, 1,
\end{align}
and for the potential outcome we have
\begin{align}\label{identification_counter_outcome}
	\mathbb{E}[I_{a,m}|A=1-a] = \mathbb{E}[\mathbb{E}[I|A=a, M=m, L]|A=1-a], \quad a = 0, 1.
\end{align}
Similar reasoning can be applied to the  total benefit~$p_t(a)$ and the direct benefit ~$p_d(a)$ in the $a$th group (as they appear in eq.~\eqref{prod_dir_plus_indir}). The estimators of $\mathbb{E}[M|A=a, L]$ and $\mathbb{E}[I|A=a, M=m, L]$ can be obtained by using standard regression models. The expectation with respect to the  conditional distribution given $A=1-a$ in equations~\eqref{identification_counter_med} and~\eqref{identification_counter_outcome} can be calculated by computing the sample average of the regression model over the~$L$ values in the $(1-a)$th exposure group.\cite{bender2007estimating}  The estimation procedure, presented in the following sections, first targets the effects themselves - the primary estimands - and then derives the novel effect-size indices by applying~$g$ (Eq.~\eqref{def.g}) to the resulting estimates.

\section{Parametric Models}

\subsection{The indirect effect indices:  INNE, IEIN, and INNT}
In order to provide an explicit functional expression of the  indirect effects, we need to model the expected value of the mediator as a function of the exposure~$A$ and the measured confounders~$L$, namely,  $\mathbb E [M|A, L]$. We assume that the link function of this model is~$\eta^{-1}$ and is indexed by a set of parameters~$\gamma$. Formally, 
\begin{align}\label{mediator_model}
 \mathbb{E}[M|A, L] = \eta \left(A, L; \gamma   \right) .		
\end{align}
Additionally, we need to model the expected outcome as a function of the exposure, the mediator and the measured confounders, namely,  $\mathbb E [I| A, M, L]$.  Assume a  model with link function~$\xi^{-1}$ that is indexed by a set of parameters~$\beta$. Formally, we assume that the conditional outcome model is 
\begin{align}\label{cond_outcome_model}
	 \mathbb{E}[I|A, M, L] =   \xi \left( A, M, L; \beta     \right) .
	 \end{align}
Therefore, using the mediation formula and Theorem~\ref{thm:NIE} for the~$a$th group,  the indirect benefit for the $a$th groups is
\begin{align}\label{indirect_prob}
	p_i(a; \theta)  = &\mathbb E [ \eta \left(1, L; \gamma   \right) - 
	\eta \left(0, L; \gamma   \right) |A=a] 
	 \mathbb E [ \xi \left( 0, 1, L; \beta  \right) - 
	\xi \left( 0, 0, L; \beta  \right) |A=a] , \quad a \in \{0, 1\},
\end{align}	
where $\theta$ denotes the set of all unknown parameters $\theta ^ T = (\gamma ^ T, \beta ^ T)$ that the indirect effect is dependent on. The INNE is $g(p_i(0; \theta))$, the IEIN is $g(p_i(1; \theta))$ and the INNT is $g(\mathbb E[p_i(A; \theta)])$. 
For detailed derivation of eq.~\eqref{indirect_prob} please refer to Appendix~\ref{app:TE_id}.

\subsection{The direct effect indices: DNNE, DEIN, and DNNT}
For the direct effect in the $a$th group we need the mediator-exposure model as in eq.~\eqref{mediator_model}, and the conditional outcome model  as defined in eq.~\eqref{cond_outcome_model}. Using the mediation formula as in eq.~\eqref{def:NIE_BINARY}, and the result of Theorem~\ref{thm:NDE}, the indirect effect is
\begin{align}\label{direct_prob}
	p_d(a; \theta)  = 	&  \mathbb E [ \xi\left( 1, 0, L ; \beta  \right)  - \xi\left( 0, 0, L ; \beta  \right)
	 |A=a] ( 1 -  \mathbb E [ \eta\left(1, L; \gamma  \right) |A=a ])\\
	& +
	 \mathbb E [ \xi\left( 1, 1, L ; \beta  \right)  - \xi\left( 0, 1, L ; \beta  \right)
	 |A=a]  \mathbb E [ \eta\left(1, L; \gamma  \right) |A=a ],\nonumber
\end{align}
for $ a \in \{0, 1\}$.	The DNNE is $g(p_d(0; \theta))$, the DEIN is $g(p_d(1; \theta))$ and the DNNT is $g(\mathbb E[p_d(A; \theta)])$. 

\subsection{The total effect indices:  NNE, EIN, and NNT}
Following the computation of the indirect and direct benefits, which are the primary parameters of interest in this type of analysis,  we can use them to compute the total effect~$p_t$. Direct computations of the total (marginal) effect indices is, in general, also possible. Assuming a link function $\xi^{-1}$ for the conditional outcome as in eq.~\eqref{cond_outcome_model}, one can obtain the marginal outcome  model by marginalizing it over~$M$. Namely, the outcome marginal model  is computed by 
\begin{align}\label{marginal_model}
	\mathbb{E}[I|A, L] =  \mathbb{E} \left[  \mathbb E [I |A, M, L]| A, L\right] =
	\mathbb{E} \left[  \xi \left( A, M, L; \beta \right) | A, L\right] . 
\end{align}
The explicit form of $\mathbb{E}[I|A, L]$ depends on the link function~$\xi^{-1}$. For linear and log link functions the additive linearity on the $\xi^{-1}$ scale  with respect to the parameters~$\beta$ is preserved in the marginal outcome model, i.e., for such link functions 
 \begin{align}\label{marginal_model-glink}
 	\mathbb{E}[I|A, L] = \xi \left( A, L; \beta^*   \right),
 \end{align}
 where the vector of parameters~$\beta^*$ are functions of the vector of parameters~$\beta$ of the conditional outcome model from eq.~\eqref{cond_outcome_model}.
 This relationship also holds  approximately for the logit link function under rare outcome scenario since in such scenario the logit link function resembles the log function.\cite{nevo2017estimation} However, for other link functions, this relationship does not hold. Therefore, assuming such marginal model as in eq.~\eqref{marginal_model-glink} can only be viewed as approximation of the true marginal outcome model and is not guaranteed to be valid. Alternatively, one can model the marginal relationship between the outcome and the exposure, given the measured covariates, and then to use the total effect decomposition in order to estimate the direct effect. In such scenario the conditional outcome model in eq.~\eqref{cond_outcome_model} can be viewed only as an approximation of the true conditional outcome model, and is not guaranteed to be valid. For more details on the identification and estimation of the marginal indices (NNT, NNE, EIN) please refer to~\cite{VancakSjölander+2024} and \cite{bender2010estimating}.

\section{Estimation and Inference}
Estimation and inference are conducted using the M-estimation method,\cite{stefanski2002calculus} also known as the estimating equations method. This approach finds estimators by solving a set of equations derived from the observed data, without requiring a specified distribution for the data. M-estimation allows for the incorporation of all sources of sampling variability, including the variability arising from using the sample distribution of~$L$ given~$A$, and the marginal distribution of~$A$.
 Since our target parameters are functions of the indirect, direct, and total effects, defined by applying the function~$g$. Let   $(\beta_I, \gamma, \mathbf{p}_i, \mathbf{p}_d, \mathbf{p}_t , 
\mathbf{g}_i, \mathbf{g}_d, \mathbf{g}_t)^T$ be the vector of all estimands, where $\mathbf{p}_i = (p_i(0), p_i(1), p_i)^T$, $\mathbf{p}_d = (p_d(0), p_d(1), p_d)^T$, and $\mathbf{p}_t = (p_t(0), p_t(1), p_t)^T$. Additionally, $\mathbf{g}_i$ is the indirect effect triple, i.e., (INNE, IEIN, INNT)$^T$, $\mathbf{g}_d$ is the direct effect triple, i.e., (DNNE, DEIN, DNNT)$^T$, and $\mathbf{g}_t$ is the total effect triple, i.e., (NNE, EIN, NNT)$^T$.  Therefore, the  vector-valued estimating function is defined as
\begin{align}\label{eq:est_fun}
	\mathbf{Q} (M, L, A; \theta ) =
	\begin{pmatrix}
		\mathbf{S}(M, L, A; \beta_I, \gamma)\\
		\Xi (M, L, A; \beta_I)\\
		\mathcal{H} (L, A; \gamma)\\
		\mathbf{p}\left(M, L, A; \beta_I, \gamma , \mathbf{p}_i \right)\\
		\mathbf{p}\left(M, L, A; \beta_I, \gamma, \mathbf{p}_d \right)\\
		\mathbf{p}\left(M, L, A; \beta_I, \gamma, \mathbf{p}_t \right)\\
		\mathbf{g}\left(\mathbf{p}_i, \mathbf{g}_i\right)\\
		\mathbf{g}\left(\mathbf{p}_d,\mathbf{g}_d\right)\\
		\mathbf{g}\left(\mathbf{p}_t, \mathbf{g}_t\right)\\
	\end{pmatrix},
\end{align}
and the corresponding estimating equations are
\begin{align}\label{eq:est_eq}
	\sum_{j=1}^n \mathbf{Q} ( M_j, L_j, A_j; \theta ) = 0.
\end{align}
The vector-valued function  $\mathbf{S}(M, L, A; \beta_I, \gamma )$ is a  vector of unbiased estimating functions of~$\beta_I$,  and~$\gamma$ 
respectively. Namely,
\begin{align}\label{eq:est_fun_b}
	(S(M, L, A; \beta_I) , S(A, L ; \gamma))^T. 
\end{align}
The vector-valued function $\Xi (M, L, A; \beta_I)$ is
\begin{align}\label{est:Xi_vec}
	\begin{pmatrix}
			\left(\xi ( 1, 0, L ;  \beta  )  - \xi ( 0, 0, L ;  \beta ) - 
	\mathbb{E}[I_{1, 0} - I_{0, 0}|A=0]  \right) (1-A)\\
	\left(\xi ( 1, 0, L ;  \beta  )  - \xi ( 0, 0, L ;  \beta ) - 
	  \mathbb{E}[I_{1, 0} - I_{0, 0}|A=1] \right) A\\
	\left(\xi ( 0, 1, L ;  \beta  )  - \xi ( 0, 0, L ; \beta ) - 
	  \mathbb{E}[I_{0, 1} - I_{0, 0}|A=0]  \right) (1-A)\\
	\left(\xi ( 0, 1, L ;  \beta  )  - \xi ( 0, 0, L ; \beta ) - 
	\mathbb{E}[I_{0, 1} - I_{0, 0}|A=1] \right) A\\
	 \left( \xi\left( 1, 1, L ; \beta  \right)  - \xi\left( 0, 1, L ; \beta  \right) -
	 \mathbb{E}[I_{1, 1} - I_{0, 1}|A=0] \right) (1-A)  \\
	 	 \left( \xi\left( 1, 1, L ; \beta  \right)  - \xi\left( 0, 1, L ; \beta  \right) -
	 \mathbb{E}[I_{1, 1} - I_{0, 1}|A=1] \right) A
\end{pmatrix}.
\end{align}
The vector-valued function 	$\mathcal{H} ( L, A; \gamma) $ is
\begin{align*}
	\begin{pmatrix}
	 ( \eta\left(1, L; \gamma  \right) - \mathbb{E}[M_1|A=0])(1-A) \\
( \eta\left(1, L;  \gamma  \right) - \mathbb{E}[M_1|A=1])A \\ 
( \eta\left(0, L;  \gamma  \right) - \mathbb{E}[M_0|A=0])(1-A) \\
( \eta\left(0, L;  \gamma  \right) - \mathbb{E}[M_0|A=1])A \
	\end{pmatrix}.
\end{align*}
The vector-valued function $ 	\mathbf{p}\left(M, L, A; \beta_I, \gamma , \mathbf{p}_i \right)$ is 
\begin{align}\label{eq:est_fun_p}
	((p_i(0; M, L,  \beta_I, \gamma) - p_i(0))(1-A),
	(p_i(1; M, L, \beta_I, \gamma) - p_i(1))A,
	p_i(0)(1-A) + p_i(1)A  - p_i  ) ^T,	
\end{align}
where $p_i(0; M, L, \beta_I,  \gamma)$ and $p_i(1; M, L, \beta_I, \gamma)$ are the conditional indirect groupwise exposure effects as defined in eq.~\eqref{indirect_prob} for $a=0$ and $a=1$, respectively, and the mean of $p_i(0)(1-A) + p_i(1)A $ is the marginal indirect exposure effect.  Finally, the vector-valued function  $ 	\mathbf{g}\left(\mathbf{p}_i, \mathbf{g}_i\right)$ is  
\begin{align}\label{eq:est_fun_g}
	(g(p_i(0)) - \text{INNE},
	g(p_i(1)) - \text{IEIN},
	g(p_i ) - \text{INNT}  ) ^T,
\end{align}
which is required for estimating the corresponding indices. Notably, although this function is independent of the observed data, it is still required for computing the asymptotic variance of the estimators of the INNE, IEIN, and INNT, respectively.  The vector-valued functions  $ 	\mathbf{p}\left(M, L, A; \beta_I, \gamma, \mathbf{p}_d \right)$,
$\mathbf{p}\left(M, L, A; \beta_I, \gamma, \mathbf{p}_t \right)$, 
and $
\mathbf{g}\left(\mathbf{p}_d,\mathbf{g}_d\right)$,
$\mathbf{g}\left(\mathbf{p}_t, \mathbf{g}_t\right)$,  are defined analogously for the direct and the total effects as defined in eq.~\eqref{direct_prob}.

\noindent The asymptotic variance of~$\theta$'s estimators~$\hat \theta$ is obtained by applying the sandwich formula
\begin{align}\label{eq:sadwitch}
	\mathbf{V}(\hat \theta) = n^{-1} \mathbf{A}(\theta)^{-1} \mathbf{B}(\theta) \mathbf{A} (\theta)^{-T}, 
\end{align} 
where the “bread” matrix is the minus Jacobian matrix of the estimating function, i.e.,  $\mathbf{A}(\theta) = \mathbb{E}[-\partial_{\theta} \mathbf Q(\theta)/\partial \theta^{T}]$, the “meat” matrix is the outer product of the  estimating function, i.e., $\mathbf{B}(\theta) = \mathbb{E}[\mathbf Q(\theta) \mathbf Q(\theta)^T]$, where $\mathbf{Q}(\theta)$ is a shorthand for the vector valued estimating function from eq.~\eqref{eq:est_fun}. The asymptotic distribution of the estimators~$\hat{\theta}$ is multivariate normal
\begin{align*}
	\sqrt{n} (\hat \theta - \theta ) \xrightarrow{D} \mathcal{N}_p(0, \mathbf{V}(\theta)), 
\end{align*}  
where the subscript~$p$ denotes the dimension of the parametric space~$\theta \in \Theta$, and the superscript~$D$ denotes convergence in distribution.  For the sample version of “bread” $\mathbf A(\theta)$ and the “meat” $\mathbf B(\theta)$ matrices, we replace the expectation operator with the corresponding sample means, and $\theta$ with ts estimator~$\hat \theta$. The resulting estimated covariance matrix $\mathbf{V}(\hat \theta)$ is then used to construct analytical confidence intervals for the novel indices based on the normal approximation. 

Notably, the construction of the estimating functions in this section relies on the factorization in Subsection~\ref{subsec:factor} and its accompanying assumptions. However, dropping this factorization has no effect on (i) the definitions of the novel indices, (ii) the identification assumptions for the NIE and NDE and the corresponding indices, or (iii) the specification of the parametric working models. The only change concerns the estimating equations: M-estimation remains applicable, but the vector-valued estimating functions must be modified to reflect, for example, conditional rather than marginal factorization. In particular, this reduces the number of estimating functions in the vector-valued function~$\Xi$ given in Eq.~\eqref{est:Xi_vec}.

\subsection{Example: logit link functions}
Assume that the outcome model and the mediator model share the same link function; to such models, we refer as double-$\xi$ models. Particularly, assume that,  $\xi^{-1}(p) \equiv \eta ^{-1}(p) = \text{logit}(p) \equiv \ln \left(\frac{p}{1-p}\right)$, where~$\xi$ is the inverse logit function which is denoted by expit$(x) = \frac{1}{1 + e^{-x}}$.  Assume that~$L$ is a univariate continuous random variable. Therefore, the mediator and the outcome models are given as
\begin{align*}
\mathbb{E}[M| A, L] &= \eta(A, L; \gamma) = \text{expit} \left(  \gamma_0 + \gamma_A A + \gamma_L L  \right), \\
\mathbb{E}[I| A, M, L] & = \xi(A, M, L; \gamma)  =  \text{expit} \left(  \beta_0 + \beta_A A + \beta_M M  + \beta_L L  \right).
\end{align*}
The vector of the unknown parameters that are needed to be estimated in order to compute the indirect, direct and total effect triples are $(\gamma_0, \gamma_A, \gamma_L, \beta_0, \beta_A, \beta_M, \beta_L)$.  Assume that there are~$n_a$ observations in the~$a$th group, $a\in\{0,1\}$, such that the overall sample size is $n = n_0 + n_1$. Therefore, the indirect effect  in the $a$th group~$p_i(a)$ can be estimated by replacing the unknown parameters with their M-estimators, replacing the expectation operator with the sample mean in the corresponding group, and then computing the multiplication as defined in eq.~\eqref{indirect_prob}, i.e.,
\begin{align}
\hat p_i(a ) \equiv p_i(a; \hat \theta)   
= & \frac{1}{n_a} \sum_{j=1}^{n_a}  \left( \text{expit}\{\hat \gamma_0 + \hat \gamma_A  + \hat \gamma_L L_j\} - \text{expit}\{\hat \gamma_0 + \hat \gamma_L L_j\}    \right) \nonumber \\  
\times 
& \frac{1}{n_a} \sum_{j=1}^{n_a}  \left( \text{expit}\{\hat \beta_0 + \hat \beta_M + \hat \beta_L L_j\} - \text{expit}\{\hat \beta_0 + \hat \beta_L L_j\}    \right) \, \nonumber .
\end{align}
The estimator of the corresponding index for the $a$th group (INNE, IEIN) is obtained by applying the function~$g$ as defined in~\eqref{def.g} to~$\hat p_i(a)$. The INNT is obtained by applying the function~$g$  to the marginal indirect~$\hat p _i$ that is obtained by the weighted mean of the indirect groupwise effects, i.e., 
\begin{align*}
\hat p_i  \equiv p_i(\hat \theta ) &=  
\frac{1}{n}\sum_{j=1}^n \left( a_j  p_i(0; \hat \theta ) + (1-a_j)  p_i (1; \hat \theta ) \right) \\
& = \frac{n_0}{n}	 p_i(0; \hat \theta) + \frac{n_1}{n} p_i(1; \hat \theta). 
\end{align*} 
For the direct effect~$p_d$, analogical steps can be applied to obtain the sample analogue of eq.~\eqref{direct_prob}, i.e., 
\begin{align*}
	\hat p_d(a) \equiv p_d(a; \hat \theta)  & 
	=  
	 \frac{1}{n_a} \sum_{j=1}^{n_a}  \left( \text{expit}\{\hat \beta_0 + \hat \beta_A + \hat \beta_L L_j\} - \text{expit}\{\hat \beta_0 + \hat \beta_L L_j\}    \right)\\ 
	&\times   
	\left( 1 - \frac{1}{n_a} \sum_{j=1}^{n_a}\text{expit}\{\hat \gamma_0 +  \hat \gamma_A + \hat \gamma_L L_j\}    \right)\\
	&  + \frac{1}{n_a} \sum_{j=1}^{n_a}  \left( \text{expit}\{\hat \beta_0 + \hat \beta_A +\hat \beta_M + \hat \beta_L L_j\} - \text{expit}\{\hat \beta_0 + \hat \beta_M +  \hat \beta_L L_j\}    \right)\\ 
	&\times   
	\frac{1}{n_a} \sum_{j=1}^{n_a} \left( \text{expit}\{\hat \gamma_0 +  \hat \gamma_A + \hat \gamma_L L_j\}    \right).
\end{align*}
The estimator of the corresponding index for the~$a$th group (DNNE, DEIN) is obtained by applying the function~$g$ as defined in~\eqref{def.g} to~$\hat p_d(a)$. The DNNT is obtained by applying the function~$g$  to the marginal indirect~$\hat p _d$ that is obtained by the weighted mean of the direct groupwise effects, i.e., 
\begin{align*}
	\hat p_d
	 = \frac{n_0}{n}	 p_d(0; \hat \theta) + \frac{n_1}{n} p_d(1; \hat \theta). 
\end{align*} 
 The estimators for the total effect triple - NNE, EIN, NNT - are derived by applying the function~$g$ as defined in equation~\eqref{def.g}, to the marginal groupwise total effect. Finally, the total effect NNT is calculated by applying the function~$g$  to the sum of the estimated marginal indirect and direct effects as defined in eq.~\eqref{prod_dir_plus_indir}
\begin{align*}
	\hat p_t =  \hat p_i +  \hat p_d.  
\end{align*}
In practice, all nine indices and their corresponding parameters are estimated simultaneously. The sequential presentation here is for didactic purposes only. The corresponding procedure for the probit link function is analogous, with the logistic inverse link function replaced by the standard normal cumulative distribution function.

\section{Simulation study}
In order to illustrate the  properties of the M-estimators of the new indices and the corresponding CIs, we consider two main simulation setups: one with the logit link function and the other with the probit link function for both the outcome and the mediator models (i.e., double logit and double probit models).We consider a binary outcome~$I$, binary exposure~$A$, binary mediator~$M$, and a continuous univariate confounder~$L$. For ease of interpretation, one may follow the numerical example in subsection~\ref{sec: num_example}, which concerns emigration patterns. In this context, the exposure~$A$ indicates possession of an undergraduate academic degree, the outcome~$I$ represents emigration, and the mediator~$M$ reflects command of a foreign language. The confounder~$L$ may represent age, that affects the probability of acquiring a degree, attaining language proficiency, and emigrating. Figure~\ref{dag:sim_mod} illustrates the DAG of the data-generating process. Further details of the simulation setup are provided in the next subsection.

\begin{center}	
\begin{figure}
	\[
	\xymatrixrowsep{0.7cm}
	\xymatrixcolsep{1cm}
	\xymatrix{ 
		A \ar@/_-1.5pc/[urrrrd] \ar[rr]&& M \ar[rr] && I   \\
		&& L \ar[llu] \ar[u] \ar[rru] 
		&&&
	}
	\]
	\caption{\small{DAG the simulated models.  $A$ is a binary exposure, $M$ is a binary mediator, $L$ is a normally distributed confounder, and~$I$ is the binary outcome.}  }\label{dag:sim_mod}
\end{figure}
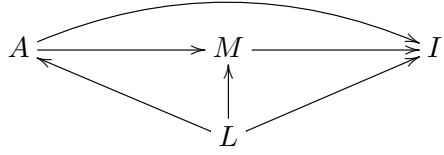
\end{center}

\subsection{Simulation setup}
We simulated data to resemble a cohort study with  binary outcome ~$I \in \{0, 1\}$, a binary mediator~$M\in \{0, 1\}$, binary exposure~$A \in \{0, 1\}$, and normally distributed univariate  confounder~$L$. In the first step, we specify the link functions of the exposure-confounder, mediator-outcome and the conditional outcome models and their parameters. In the second step, we generate the data. In the third step, we use the data to estimate the  the indices of interest and their corresponding 95$\%$-level analytical confidence intervals. In the fourth step, we assess the efficiency and accuracy of our method by evaluating the coverage rates of the  $95\%$-level sandwich matrix-based CIs.  The structural models are

\begin{align}
	L &\sim N(\mu, \sigma^2),\\
	A|L& \sim Ber\left(   \text{expit}\left( \delta_0 + \delta_L L\right)  \right) \nonumber \\
	 M | A, L &\sim Ber\left(   \xi \left( \gamma_0 + \gamma_AA +  \gamma_LL\right)  \right) \nonumber \\
	 	 I | A, M, L &\sim Ber\left(   \xi \left( \beta_0 + \beta_AA + \beta_M M + \beta_LL\right)  \right) \nonumber. 
\end{align}
For both examples, the parameters of the structural models were set to $\mu = 0.5$, $\sigma = 0.1$, $\delta = (2, -3)^T$, $\gamma = (-1, 3, -2)^T$, and $\beta = (-1, 1.5, 1.5, -2)^T$. Under the logit link function, defined as $\xi(x) = \text{expit}(x) = \frac{1}{1 + e^{-x}}$, the true values of the nine indices were: DEIN = 3.07, DNNE = 3.08, DNNT = 3.07; IEIN = 6.28, INNE = 6.53, INNT = 6.37; EIN = 2.06, NNE = 2.09, and NNT = 2.07. For the probit link function, defined as $\xi(x) = \text{probit}(x) = \Phi(x)$, the corresponding true values were: DEIN = 2.06, DNNE = 2.06, DNNT = 2.06; IEIN = 4.18, INNE = 4.49, INNT = 4.29; and EIN = 1.38, NNE = 1.41, NNT = 1.39. All true values were computed using Monte Carlo integration.

The system of estimating equations consisted of 32 equations corresponding to 32 unknown parameters, including the nine proposed indices. This system was solved numerically, and the resulting point estimates were used to compute the asymptotic covariance matrix via the sandwich formula. The simulation study was conducted for sample sizes $n = 200, 400, 800, 1600$, with $k = 100$ distinct data sets generated for each sample size. 
 The full simulation source code, along with the generated point estimates and corresponding confidence interval limits, is available in the first author GitHub repository.\footnote{Simulations source code: 
 }

\subsection{Simulation results and summary}
\begin{center}
	\begin{figure}[ht]
		\centering
		\includegraphics[width=1\textwidth, height=.6\textwidth]{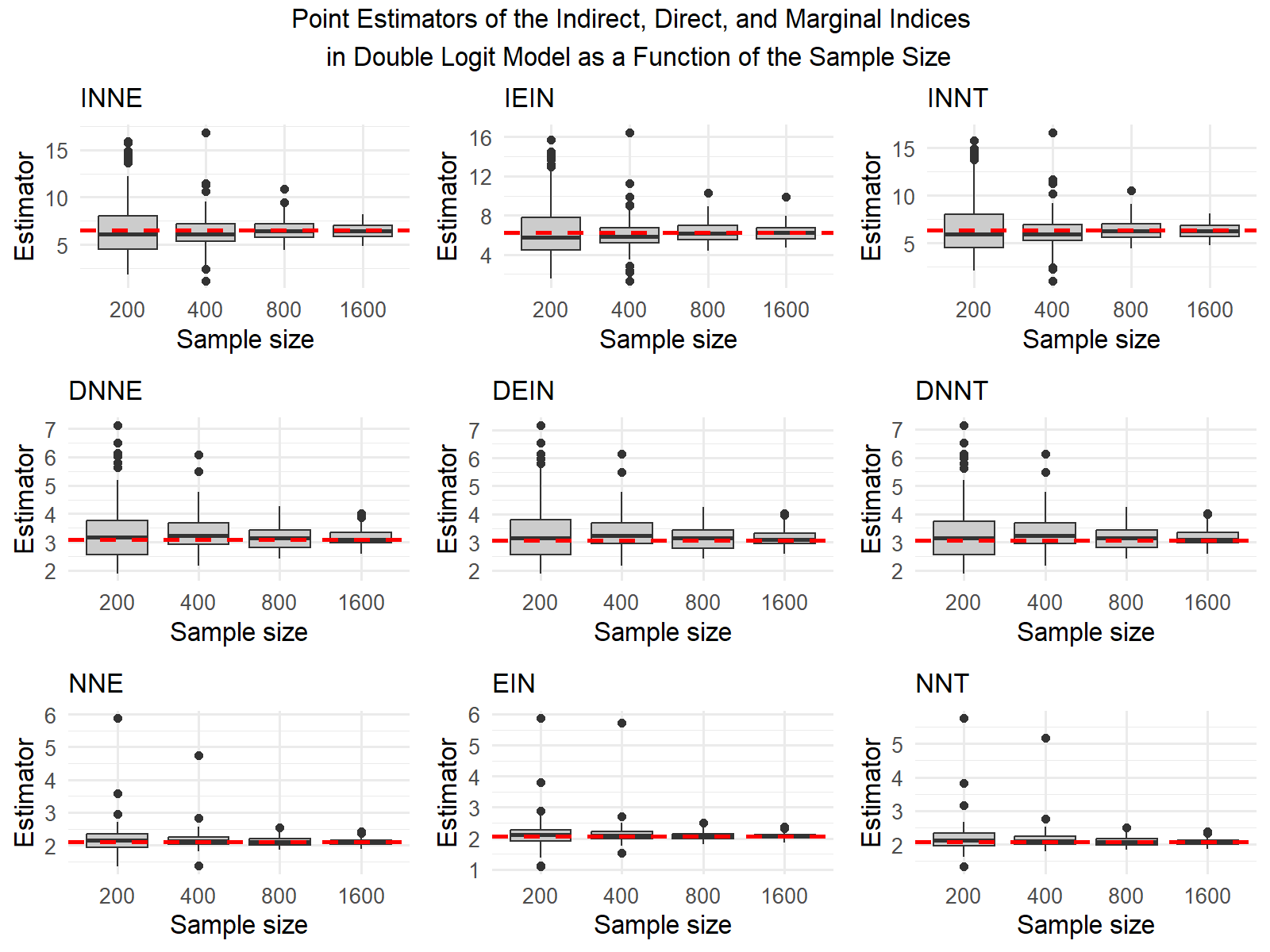}
		\caption{\small{Double logit model simulation: Boxplots of the M-estimators of the nine effect indices: INNE, IEIN, INNT (indirect effects); DEIN, DNNE, DNNT (direct effects); and NNE, EIN, NNT (marginal effects). The red dashed lines denote the true parameter values: INNE = 6.53, IEIN = 6.28, INNT = 6.37; DEIN = 3.07, DNNE = 3.08, DNNT = 3.07; NNE = 2.09, EIN = 2.06, NNT = 2.07. The simulated sample sizes are~$n = 200, 400, 800, 1600$, and the number of iterations for each sample size is $k = 100$ (extremely large values exceeding three times the true value were omitted for clarity). }}\label{fig:sim_direct_logit}
	\end{figure}
\end{center}

\begin{center}
	\begin{figure}[ht]
		\centering
		\includegraphics[width=1\textwidth, height=.6\textwidth]{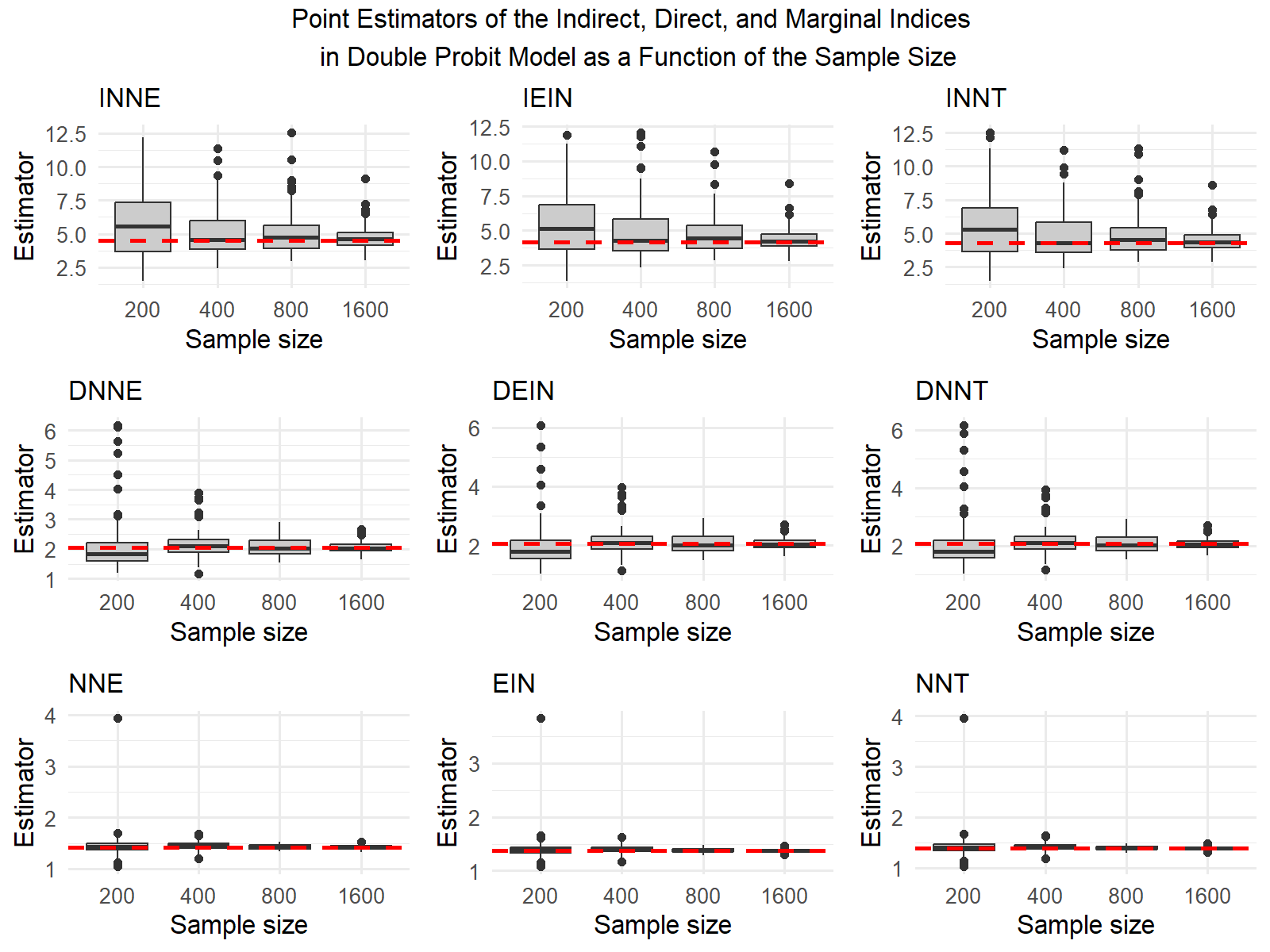}
		\caption{\small{Double probit model simulation: Boxplots of the M-estimators of the nine effect indices: INNE, IEIN, INNT (indirect effects); DEIN, DNNE, DNNT (direct effects); and NNE, EIN, NNT (marginal effects). The red dashed lines denote the true parameter values: INNE = 4.49, IEIN = 4.18, INNT = 4.29; DEIN = 2.06, DNNE = 2.06, DNNT = 2.06; NNE = 1.41, EIN = 1.38, NNT = 1.39. The simulated sample sizes are~$n = 200, 400, 800, 1600$, and the number of iterations for each sample size is $k = 100$ (extremely large values exceeding three times the true value were omitted for clarity).
		 }}\label{fig:sim_direct_probit}
	\end{figure}
\end{center}

A graphical summary of the estimators’ behavior as a function of the sample size~$n$ for the double logit and the double probit models can be found in Figure~\ref{fig:sim_direct_logit} and Figure~\ref{fig:sim_direct_probit}, respectively. The boxplots illustrate the stability and the consistency of the M-estimators of the nine indices for the two models. For  numerical summary, please refer to Table~\ref{tb:logit_coverage_rate} that presents the empirical coverage rates of the 95\%-level  analytical sandwich-based CIs for the nine indices for each one of the models as a function of the sample size~$n$. 

The relatively small sample sizes were intentionally chosen to evaluate the performance of the proposed method - derived from the asymptotic theory - under more modest sample size conditions. While efficient performance is expected in large cohort studies, this is less informative and is partially demonstrated in our simulations for the largest sample size ($n = 1600$). Notably, the empirical coverage rates approached the nominal 95\% level as the sample size increased. For the double logit model, near-nominal coverage was already achieved at $n = 400$, whereas the double probit model required a larger sample size of $n = 1600$ to attain comparable performance. At $n = 200$, the double probit model exhibited substantially lower coverage for four of the nine indices, while the double logit model showed relatively stable coverage even at this smaller sample size, with only modest deviations from the nominal level. Nevertheless, for $n = 200$, approximately 7\% of the iterations in the logit model resulted in either infinite confidence intervals or singular covariance matrices, compared to only about 2\% in the probit model. This behavior is not unexpected and does not reflect any inherent theoretical flaw. Rather, it stems from the high dimensions of the parametric space, which involves solving a system of 32 equations for 32 parameters and computing a $32 \times 32$ covariance matrix. Naturally, such a high-dimensional estimation problem requires a sufficiently large sample size to ensure numerical stability. Indeed, for both models, these issues were no longer observed at sample sizes of $n = 800$ and above.

\begin{table}[htbp]
	\centering
	\begin{tabular}{c|cccc|cccc}
		&& \text{logit} && && \text{probit} &&\\
		\hline
		$n$ & 200 & 400 & 800 & 1600 & 200 & 400 & 800 & 1600  \\ 
		\hline
		INNE & 0.85 & 0.90 & 0.97 & 0.93 & 0.80 & 0.90 & 0.96 & 0.94 \\ 
		IEIN & 0.79 & 0.91 & 0.97 & 0.92 & 0.80 & 0.91 & 0.97 & 0.94 \\ 
		INNT & 0.82 & 0.91 & 0.97 & 0.93 & 0.79 & 0.90 & 0.96 & 0.94 \\
		DNNE & 0.89 & 0.95 & 0.94 & 0.99 & 0.70 & 0.89 & 0.87 & 0.94 \\ 
		DEIN & 0.88 & 0.95 & 0.94 & 0.99 & 0.65 & 0.89 & 0.87 & 0.94 \\ 
		DNNT & 0.88 & 0.95 & 0.94 & 0.99 & 0.65 & 0.89 & 0.87 & 0.94 \\
		NNE  & 0.90 & 0.95 & 0.95 & 0.97 & 0.88 & 0.93 & 0.99 & 0.94 \\ 
		EIN  & 0.88 & 0.93 & 0.95 & 0.97 & 0.92 & 0.90 & 0.97 & 0.95 \\ 
		NNT  & 0.91 & 0.94 & 0.95 & 0.97 & 0.91 & 0.92 & 0.98 & 0.94 \\ 
		\hline
		$\%$  Inf. CIs & 7.66$\%$ & 2.11$\%$& 0$\%$& 0$\%$ &2.55$\%$ & 0.33$\%$ &  0$\%$ & 0$\%$ \\
	\end{tabular}
	\caption{\small{Empirical coverage rates of the analytical $95\%$ confidence intervals for the nine indices, under the double logit and double probit models, for sample sizes~$n = 200, 400, 800, 1600$. Each sample size corresponds to $k = 100$ simulation iterations. Coverage was calculated conditional on finite point estimates and a non-singular estimated covariance matrix. The bottom row reports the percentage of iterations with either an infinite point estimate or a singular covariance matrix.
	}}
	\label{tb:logit_coverage_rate}
\end{table}

\section{Summary and discussion}
The NNT is an efficacy and effect size measure commonly used in epidemiological studies and meta-analyses. The NNT was originally defined as the average number of patients needed to be treated to observe one less adverse effect. In this study, we introduce the novel path-dependent indices. Particularly, the direct and indirect number needed to treat (DNNT and INNT, respectively). The DNNT and the INNT are efficacy measures defined as the average number of patients that need to be treated to benefit from the treatment's direct and indirect effects, respectively. We defined these measures using nested potential outcomes. Next, we formulated the conditions for the identification of the DNNT and INNT, as well as for the direct and indirect number needed to expose (DNNE and INNE, respectively) and the direct and indirect exposure impact number (DEIN and IEIN, respectively) in observational studies. Next,  we presented an estimation method with two analytical examples. A corresponding simulation study followed these examples. The simulation study illustrated that the estimators of the novel indices are consistent, and their asymptotically-correct $95\%$-level confidence intervals meet the nominal coverage rates. 

The novel indices build on the same idea as the original NNT, NNE, and EIN. They change the effect scale while preserving information by applying a monotone transformation that translates abstract probabilities into expected numbers of individuals. Section 3 explains the assumptions under which this interpretation is especially natural and how the direct and indirect NNTs align with the original motivation for NNT-type measures. In those probabilistic settings, the novel indices are particularly useful because they express path-specific benefits as the average number of individuals who benefit through the direct or mediated path, which makes communication and decision-making more straightforward. By contrast, the mediation proportion is different. It is a composite ratio that depends on the effect scale and the chosen decomposition of the total effect. It typically requires the effects to have the same direction and does not allow recovery of the underlying indirect and total effects from the ratio alone. When the direct and indirect effects do not share the same direction, the mediation proportion is not interpretable, yet DNNT and INNT still provide path-specific summaries. The proposed path-specific NNTs are stand-alone measures. Each is tied to a single primitive effect, such as the natural direct or indirect effect, and can be reported without requiring other components, the same direction of effects, or a particular scale. However, the novel indices are not magic numbers, and their limitations are discussed below.

Although we introduced nine indices, in practice only a subset is typically used in any given study. Researchers rarely compute effect sizes for every combination of causal paths and subpopulations because it is uncommon for all nine indices to be relevant in a single analysis. As shown in the literature review, studies typically analyze a single triple, most often one of NNE, EIN, or NNT. Alternatively, some studies may require a targeted subset. For example, a researcher may focus on all indirect effects to assess mediated pathways. Notably, restricting attention to a subset of indices also simplifies computation. It lowers the dimension of the vector-valued estimating function and, consequently, the associated covariance matrix, which can improve numerical stability and precision of variance estimates. Therefore, the introduced indices constitute more of a conceptual framework rather then practical example for a specific application.  

This study's main contribution is the introduction of concepts and the analytical framework of path-dependent NNT, NNE, and EIN. However, this study is not without limitations. In epidemiologic applications based on observational exposure data, the plausibility of sequential ignorability may be limited, as the measured covariates~$L$ may be insufficient to control exposure-outcome confounding. That is, treatment ignorability may fail. Moreover, the mediator is typically not randomized, so mediator ignorability may also fail. The credibility of the identification assumptions can be strengthened (though not guaranteed) by rich adjustment for pre-treatment covariates informed by subject-matter knowledge, correctly specified DAGs, and avoidance of post-treatment controls. Nevertheless, sensitivity analyses of the novel indices to violations of either assumption set could be pursued as future work. Such analyses can follow the parametric framework of Imai et al.~\cite{imai2010identification} and the bias-formula approach of VanderWeele~\cite{vanderweele2010bias}.

 Additionally, we discuss only the direct and indirect measures, while the concept of path-dependent efficacy and size effect measures is not limited to these two causal paths. Future research prospective may include a definition and estimation of effects size that quantify more complex path dependencies. For example, indices that suit situations with multiple dependent mediators. An additional limitation is the consideration of only binary exposures and mediators. While the conceptual extensions to non-binary variables are straightforward, this extension may still introduce technical and computational challenges. With a continuous mediator, some elements carry over directly while others require additional structure. Sequential ignorability keeps the same form. Positivity must be expressed as overlap of the conditional density of the mediator, and sums over mediator categories are replaced by integrals (e.g., as in Proposition~2). Factorizations that rely on a finite set of principal strata (e.g., Proposition~1) do not apply as written. Moving to a continuous mediator requires either coarsening the mediator or imposing parametric structure on the joint distribution of the mediator under the two exposure levels. Under sequential ignorability, estimation of the natural direct and indirect effects proceeds analogously by modeling the outcome regression and the mediator mechanism and then obtaining the effects via g-computation or related semiparametric estimators.\cite{imai2010identification,valeri2013mediation}
  
  Additionally, the analytical results in this study are based on asymptotic theory. This applies both to the point estimators and to the coverage rates of the confidence intervals, as demonstrated in the simulation study. Specifically, all point estimators exhibit some degree of finite-sample bias due to the properties of the function~$g$ defined in~\eqref{def.g}, though this bias vanishes asymptotically. Similarly, the empirical coverage rates of the confidence intervals approach the nominal level as the sample size increases. While analytical derivations provide a rigorous theoretical foundation, caution is advised when applying these methods to very small sample sizes, where the asymptotic approximations may not be sufficiently accurate.   
Another natural research direction is to relax the identification assumptions. This may include relaxing the assumptions of no unmeasured confounding between the mediator and the outcome or between the exposure and the mediator. More broadly, this involves developing methods for estimating direct and indirect NNTs (NNEs, EINs) in observational studies where the set of measured covariates may not suffice to control for confounding.

The NNT, NNE, EIN, and the proposed direct and indirect indices should be interpreted with caution. These are one-dimensional summary measures that convey specific aspects of the treatment effect, rather than universal quantities capable of capturing the full complexity of the underlying causal structure.
  Their definition depends on the assumed causal structure, identification assumptions, and the fitted model. Therefore, the calculations may be sensitive to model misspecification.  However, the popularity of the NNT and the NNE in various domains, with the evident requirement for a more subtle path-dependent index, demonstrates the potential usefulness of the original and the novel direct and indirect efficacy measures. Such measures may help to distinguish and choose between direct and indirect intervention strategies and thus optimize decision-making in many practical domains, including, but not limited to, public health, migration policies, and medicine

 

\appendix

\section{APPENDIX}
\subsection{Twin causal networks}
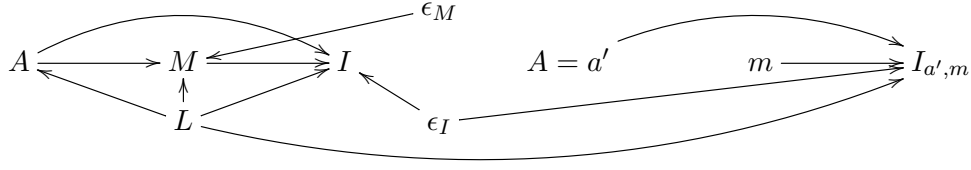
\begin{figure}
	\[
	\xymatrixrowsep{0.2cm}
	\xymatrixcolsep{0.7cm}
	\xymatrix{ &&  &&& \ar[dlll] \epsilon_M  &&&&&  &\\
		A \ar@/_-1.5pc/[urrrrd] \ar[rr]&& M \ar[rr] && I &&  A = a'  \ar@/_-1.5pc/[urrrrd]  && m \ar[rr] && I_{a', m}\\
		&& L \ar[llu] \ar[u] \ar[rru]  \ar@/_2pc/[rrrrrrrru] && &  \epsilon_I \ar[lu]  \ar[rrrrru]\\
		&&&&&
	}
	\]
	\caption{\small{DAG of a twin network causal model with a mediator that illustrates the first part of the sequential ignorability~\eqref{def:seq_ign}, namely, the conditional independence of $I_{a', m}$ and $A$, given~$L$;  $I_{a', m} \indep A | L$. The left-hand side of the DAG represents the observed actual world, while the right-hand side represents the hypothetical potential world. On the left-hand side, $M$ is the mediator, $A$ is the exposure, and~$I$ is the outcome. On the right-hand side, the exposure~$A$ is set to~$a$, the mediator~$M$ is set to~$m$,  and $I_{a', m}$ is the potential outcome where the exposure is set to $a'$, and the Mediator to~$m$.  For both DAGs, $L$ is the set of measured confounders, $\epsilon_M$ and~$\epsilon_I$ represent all the unmeasured exogenous factors that determine the values of~$M$ and~$I$, respectively.}  }\label{dag:twin_med1}
\end{figure}

\begin{figure}[ht]
	\[
	\xymatrixrowsep{0.2cm}
	\xymatrixcolsep{0.7cm}
	\xymatrix{ &&  &&& \ar[dlll] \epsilon_M  &&&&&  &\\
		A = a \ar@/_-1.5pc/[urrrrd] \ar[rr]&& M_a \ar[rr] && I_{a, M_a} &&  A = a' \ar@/_-1.5pc/[urrrrd] && m \ar[rr] && I_{a', m}\\
		&& L \ar[llu] \ar[u] \ar[rru]  \ar@/_2pc/[rrrrrrrru] && &  \epsilon_I \ar[lu]  \ar[rrrrru]\\
		&&&&&
	}
	\]
	\caption{\small{DAG of a twin network causal model with a mediator that illustrates the second part of the sequential ignorability~\eqref{def:seq_ign}, namely, the conditional independence of $I_{a', m}$ and $M_a$, given~$L$ and~$A=a$;  $I_{a', m} \indep M_a | L, A=a$. The left-hand side of the DAG represents the hypothetical world where only the exposure is set to~$a$, while the right-hand side represents the hypothetical world where both the exposure and the mediator are set to~$a$ and $m$, respectively. On the left-hand side, $M_a$ is the potential mediator where $A$ is set to~$a$,  and~$I_{a, M_a}$ is the nested potential outcome for $a$ and $M_a$. On the right-hand side, the exposure $A$ is set to $a$, and mediator~$M$ is set to~$m$, thus $I_{a', m}$ is the potential outcome for $a' \neq a$, and~$m$.  For both DAGs,~$L$ represented the set of all measured confounders, while $\epsilon_M$ and~$\epsilon_I$ represent all the unmeasured exogenous factors that determine the values of~$M$ and~$I$, respectively.}  }\label{dag:twin_med3}
\end{figure}
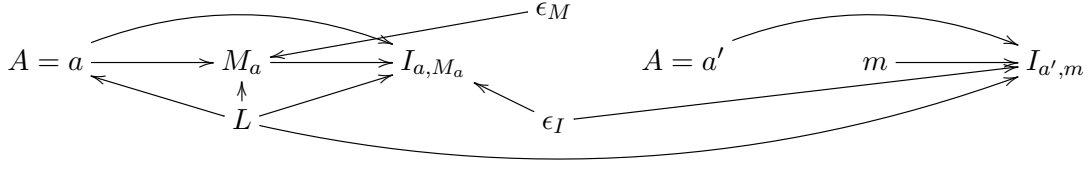

A twin network is a graphical method that presents two networks together - one for the hypothetical world and the other for the factual world or two networks for two distinct hypothetical worlds.\cite{balke2022probabilistic} Such networks provide a graphical way to check and test independence between factual and counterfactual quantities. The DAGs in Figures~\ref{dag:twin_med1} and~\ref{dag:twin_med2} illustrate the first part of the sequential ignorability assumption, i.e., the conditional independence of $\{ I_{a', m}, M_a\}$ and~$A$, given~$L=l$; $\{ I_{a', m}, M_a\} \indep A | L=l$. The DAG  in Figure~\ref{dag:twin_med3} illustrates the second part of the sequential ignorability, i.e., the conditional independence of $I_{a', m}$ and $M_a$, given~$L=l$ and~$A=a$;  $I_{a', m} \indep M_a | L=l, A=a$. For further information on Twin networks, please refer to Chapter~7 in \cite{pearl2009causality}.

\begin{figure}[ht]
	\[
	\xymatrixrowsep{0.2cm}
	\xymatrixcolsep{0.7cm}
	\xymatrix{ &&  &&& \ar[dlll] \epsilon_M \ar[drrr] &&&&&  &\\
		A \ar@/_-1.5pc/[urrrrd] \ar[rr]&& M \ar[rr] && I &&  A = a \ar@/_-1.5pc/[urrrrd] \ar[rr]&& M_a \ar[rr] && I_{a, M_a}\\
		&& L \ar[llu] \ar[u] \ar[rru] \ar@/_1.5pc/[rrrrrru] \ar@/_2pc/[rrrrrrrru] && &  \epsilon_I \ar[lu]  \ar[rrrrru]\\
		&&&&&
	}
	\]
	\caption{\small{DAG of a twin network causal model with a mediator that illustrates the first part of the sequential ignorability~\eqref{def:seq_ign}, namely, the conditional independence of $M_a$ and $A$, given~$L$;  $M_a \indep A | L$. The left-hand side of the DAG represents the observed actual world, while the right-hand side represents the hypothetical potential world. On the left-hand side, $M$ is the mediator, $A$ is the exposure, and~$I$ is the outcome. On the right-hand side, $M_a$ is the potential value of the mediator where~$A$ is set to~$a$. In addition, $A=a$ is the specified value of the exposure, and $I_{a, M_a}$ is the potential outcome where the exposure is set to $a$, and the Mediator attains the value if would have attain for~$A=a$.  For both DAGs, $L$ represents the set of measured confounders, where~$\epsilon_M$ and~$\epsilon_I$ represent all the unmeasured exogenous factors that determine the values of~$M$ and~$I$, respectively.}  }\label{dag:twin_med2}
\end{figure}

\subsection{Controlled direct effect NNT}\label{app:CDE}
\begin{defn}
	The marginal controlled direct effect (CDE) is defined as
	\begin{align}
		p_c(m) = \mathbb{E}[I_{1,m} - I_{0,m}], \quad m \in \mathcal{M},
	\end{align}
\end{defn}

\noindent where $\mathcal{M}$ is the support set of the mediator~$M$. Notably, setting~$M$ to a fixed value~$m$ removes the path $A \to M$ in Figure~\ref{dag:sim_mod}, thereby isolating the direct effect of~$A$ on~$I$ through the path $A \to I$, conditional on $M=m$. To identify $p_c(m)$ from observed data, we apply the law of total expectation:
\begin{align*}
	p_c(m) = \mathbb{E}\left[\mathbb{E}[I_{1,m} - I_{0,m} \mid M]\right].
\end{align*}
However, since $m$ is fixed, the inner conditional expectation $\mathbb{E}[I_{a,m} \mid M]$ is no longer a function of the random variable~$M$, and the outer expectation becomes redundant. That is,
\begin{align*}
	p_c(m) = \mathbb{E}[I_{1,m}] - \mathbb{E}[I_{0,m}].
\end{align*}
To express this difference in terms of observed data, we use the consistency assumption: $I = I_{a,m}$ when $A=a$ and $M=m$, and the conditional ignorability assumption, i.e.,  $I_a \indep A \mid M$ (or $I_a \indep A \mid M, L$ if covariates~$L$ are present, as illustrated in Figure~\ref{dag:twin_med1}). These yield the following identification formulas
\begin{align*}
	\mathbb{E}[I_{1,m}] = \mathbb{E}[I \mid A=1, M=m], \qquad
	\mathbb{E}[I_{0,m}] = \mathbb{E}[I \mid A=0, M=m],
\end{align*}
and therefore:
\begin{align*}
	p_c(m) = \mathbb{E}[I \mid A=1, M=m] - \mathbb{E}[I \mid A=0, M=m].
\end{align*}
These quantities can be estimated using standard regression techniques. Applying the function~$g$ to $p_c(m)$ yields the controlled direct effect on the NNT scale, also interpretable as the NNT conditional on $M=m$. An analogous definition holds for the group-specific controlled direct effect:
\begin{align*}
	p_c(m;a) = \mathbb{E}[I_{1,m} - I_{0,m} \mid A=a], \quad a = 0, 1.
\end{align*}
For additional details and applications of this framework, see~\cite{vancak2022number} and~\cite{VancakSjölander+2024}. As a final note, unlike other effect types, there is no corresponding “controlled indirect effect,” so we do not pursue this decomposition further.

\subsection{Proof of Proposition~\ref{thm:NIE}}
\begin{proof}[Proof]\label{app:proof_NIE}
	The NIE is defined as $\mathbb E[I_{a, M_1} - I_{a, M_0}]$, for a certain value~$a$ of the exposure~$A$. Without loss of generality, we assume~$a=0$ for consistency with the INNT (INNE, IEIN) defined in Definition~\ref{defn:INNT}.  Assume a binary outcome~$I$, and a binary mediator~$M$, we have 
	\begin{align*}
		\mathbb E[I_{0, M_1} - I_{0, M_0}] & = \mathbb E\left[ \mathbb  E[I_{0, M_1} - I_{0, M_0}|M_0,  M_1]\right]\\ 
		& = \mathbb E[I_{0, 1} - I_{0, 0}|M_0 = 0, M_1 = 1 ] \mathbb{P}(M_0 = 0, M_1 = 1)\\
		& + 
		\mathbb E[I_{0, 0} - I_{0, 1}|M_0 = 1, M_1 = 0] \mathbb{P}(M_0 = 1, M_1 = 0)  \\
		& = \mathbb E[I_{0, 1} - I_{0, 0}] \left( \mathbb{P}(M_0 = 0, M_1 = 1) - 
		\mathbb{P}(M_0 = 1, M_1 = 0) \right)\\
		& = \mathbb E[I_{0, 1} - I_{0, 0}]  \mathbb E[M_1 - M_0]. 
	\end{align*}
	The first equality applies the law of total expectation.  
The second equality uses the fact that for $M_1 = M_0$, the term $\mathbb{E}[I_{0, M_1} - I_{0, M_0}]$ vanishes.  The third equality stems from effect homogeneity over principal strata, i.e., 
$
\mathbb{E}\big[I_{0,1}-I_{0,0}\mid M_0, M_1\big]
= \mathbb{E}\big[I_{0,1}-I_{0,0}\big].
$
The final equality follows from the following identity
	\begin{align*}
		\mathbb E[M_1 - M_0] & = \mathbb P ( M_0 = 0, M_1 = 1)  -   \mathbb P ( M_0 = 1, M_1 = 0)  + 0 \times  
		P(M_0 = M_1)
		\\\
		& =  \mathbb{P}(M_0 = 0, M_1 = 1) - 
		\mathbb{P}(M_0 = 1, M_1 = 0) , 
	\end{align*}
	which concludes the proof.
\end{proof}

\subsection{Proof of Proposition~\ref{thm:NDE}}
\begin{proof}[Proof]\label{app:proof_NDE} 
	The NDE is defined as $\mathbb E[I_{1, M_a} - I_{0, M_a}]$, where the exposure~$A$ is set to~$a$ for the potential mediator~$M_a$. Without loss of generality, we assume~$a=1$ for consistency with the INNT (INNE, IEIN) as defined in~\ref{defn:INNT}.  Assume a binary outcome~$I$, and a binary mediator~$M$. Therefore, the NDE in the $a$th group can be written as follows 
	\begin{align*}
	p_d(a) & = 
	\mathbb{E} \left[I_{1, M_1} - I_{0, M_1} |A = a \right]\\
	& = \mathbb E [ \mathbb{E}\left[I_{1, M_1} - I_{0, M_1}|  A = a, M_1 \right] | A = a] \\
	& = \mathbb{E}\left[I_{1, 0} - I_{0, 0} | A = a, M_1 = 0 \right] ( 1 - \mathbb E [ M_1 |  A=a])
	 + 
	 \mathbb{E}\left[I_{1, 1} - I_{0, 1} | A = a, M_1 = 1 \right] \mathbb  E [ M_1 | A = a]\\
	 & =  \mathbb{E}\left[I_{1, 0} - I_{0, 0} | A = a \right] ( 1 - \mathbb E [ M_1 |  A=a])
	 + 
	 \mathbb{E}\left[I_{1, 1} - I_{0, 1} | A = a \right] \mathbb  E [ M_1 | A = a] ,
\end{align*}
where the last equality assumes 
$
\mathbb{E}[I_{1,m}-I_{0,m}\mid A=a, M_a]
= \mathbb{E}[I_{1,m}-I_{0,m}\mid A=a],
$
 for $m = 0, 1$.

\end{proof}

\subsection{Identifiability of indirect effect in the $a$th group}\label{app:TE_id}
Although the illustration below focuses on the unexposed group $A=0$, the derivation for the exposed group $A=1$ is entirely analogous, with all expectations are conditioned on $A=1$ instead.
For the INNE, which is defined as $g(p_i(0))$, we show the identifiability procedure for the first multiplicative term of eq.~\eqref{def:NIE_BINARY}. Therefore, we need to express $\mathbb E [M _1 -  M_0  | A=0]$ as a function of the observed data and the fitted model for the unexposed group~$A=0$.    Assume the exposure-mediator model as in eq.~\eqref{mediator_model}, thus
\begin{align*}
	\mathbb E [M _1 -  M_0  | A=0]	&= 
	\mathbb E [M_1 | A = 0] - E[M_0|A=0]\\
	& = 
	\mathbb E [ \mathbb E [ M | A = 1, L] | A=0] - \mathbb E [  \mathbb E[M|A=0, L] | A=0] \\
	& = 
	\mathbb E [  \eta\left( 1, L; \gamma  \right)  | A=0] - 
	\mathbb E [ \eta\left( 0, L; \gamma  \right) ) |A=0] \\
	& = \mathbb E [ \eta\left( 1, L; \gamma  \right) - 
	\eta\left( 0, L; \gamma  \right) |A=0]. 
\end{align*}
The derivation for the second multiplicative term in eq.~\eqref{def:NIE_BINARY} follows the same steps for the conditional outcome model. Namely, assuming conditional outcome model as in eq.~\eqref{cond_outcome_model}, $\mathbb{E}[I_{0,1} - I_{0,0}|A=0] $  is identified and computed as
\begin{align*}
\mathbb E [
	\xi\left(    0, 1, L; \beta \right)
	 -
	 \xi\left(    0, 0, L; \beta \right) |A=0] . 	
\end{align*}
Therefore,  the indirect benefit for the unexposed group~$p_i(a; \theta)$ is identified by
\begin{align}
	p_i(0; \theta)  = &\mathbb E [ \eta \left(1, L; \gamma   \right) - 
	\eta \left(0, L; \gamma   \right) |A=0] 
	\mathbb E [ \xi \left( 0, 1, L; \beta  \right) - 
	\xi \left( 0, 0, L; \beta  \right) |A=0] ,
\end{align}	
where $\theta$ denotes the set of all unknown parameters $\theta ^ T = (\gamma ^ T, \beta ^ T)$ that the indirect effect is dependent on. Finally, the INNE is identified by $g(p_i(0; \theta))$. Replacing the conditioning set with $A=1$, that is, evaluating the expectations with respect to the exposed group, yields the identification formula for the indirect benefit among the exposed group~$p_i(1; \theta)$, and consequently for the IEIN, defined as $g(p_i(1; \theta))$.

\bibliographystyle{unsrtnat}
\bibliography{INDIRECT_NNE}

\end{document}